\documentclass[10pt,twocolumn,letterpaper]{article}

\usepackage{cvpr}              

\usepackage[accsupp]{axessibility}  
\usepackage{amsmath}
\usepackage{amssymb}
\usepackage{array}
\usepackage{booktabs}
\usepackage{graphicx}
\usepackage{multirow}
\usepackage{xcolor}

\newcommand{\nada}[1]{}

\usepackage[pagebackref,breaklinks,colorlinks]{hyperref}

\usepackage[capitalize]{cleveref}
\crefname{section}{Sec.}{Secs.}
\Crefname{section}{Section}{Sections}
\Crefname{table}{Table}{Tables}
\crefname{table}{Tab.}{Tabs.}

\def\cvprPaperID{38} 
\def\confName{CVPR}
\def\confYear{2022}

\begin{document}

\title{
Self-supervision versus synthetic datasets:\\
which is the lesser evil in the context of video denoising?}

\author{Valéry Dewil \quad Arnaud Barral \quad Gabriele Facciolo \quad Pablo Arias\\
Université Paris-Saclay, CNRS, ENS Paris-Saclay, Centre Borelli, 91190, Gif-sur-Yvette, France\\
{\href{https://centreborelli.github.io/VDU2020-the-lesser-evil}{\tt\small https://centreborelli.github.io/VDU2022-the-lesser-evil}}
}

\maketitle

\begin{abstract}
Supervised training has led to state-of-the-art results in image and video denoising. However, its application to real data is limited since it requires large datasets of noisy-clean pairs that are difficult to obtain. For this reason, networks are often trained on realistic synthetic data. More recently, some self-supervised frameworks have been proposed for training such denoising networks directly on the noisy data without requiring ground truth. On synthetic denoising problems supervised training outperforms self-supervised approaches, however in recent years the gap has become narrower, especially for video.
In this paper, we propose a study aiming to determine which is the best approach to train denoising networks for real raw videos: \emph{supervision on synthetic realistic data} or \emph{self-supervision on real data}.
A complete study with quantitative results in case of natural videos with real motion is impossible since no dataset with clean-noisy pairs exists. We address this issue by considering three independent experiments in which we compare the two frameworks.
We found that self-supervision on the real data outperforms supervision on synthetic data, and that in normal illumination conditions the drop in performance is due to the synthetic ground truth generation, not the noise model.
\end{abstract}

\begin{figure*}[!t]
    \centering
        \subfloat{\includegraphics[width=0.249\textwidth]{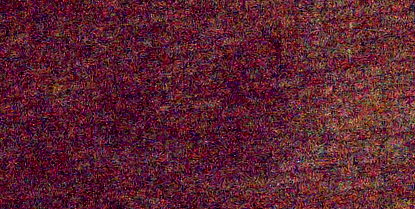}} \hfill
        \subfloat{\includegraphics[width=0.249\textwidth]{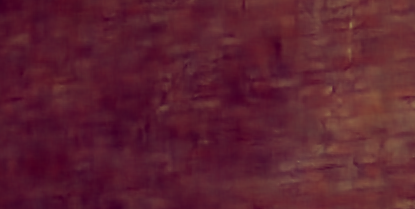}} \hfill
        \subfloat{\includegraphics[width=0.249\textwidth]{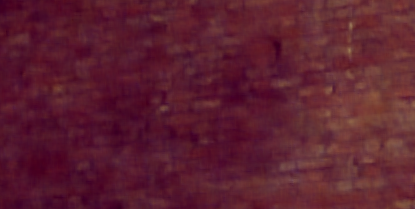}} \hfill
        \subfloat{\includegraphics[width=.249\textwidth]{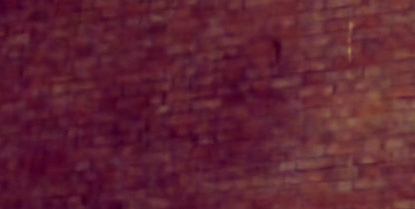}}\\
        \subfloat{\includegraphics[width=0.249\textwidth]{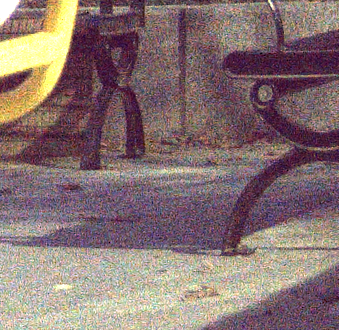}}  \hfill
        \subfloat{\includegraphics[width=0.249\textwidth]{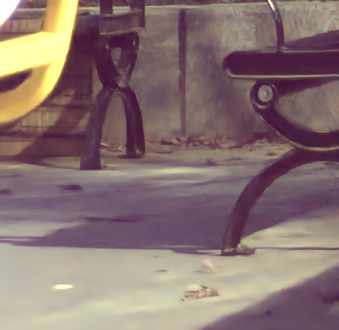}}  \hfill
        \subfloat{\includegraphics[width=0.249\textwidth]{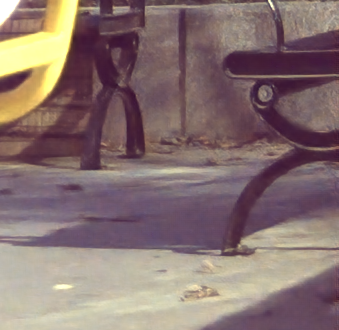}}  \hfill
        \subfloat{\includegraphics[width=0.249\textwidth]{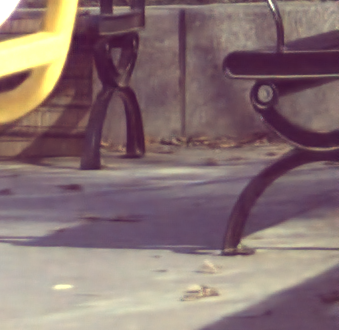}}\\
        \subfloat{\includegraphics[width=0.249\textwidth]{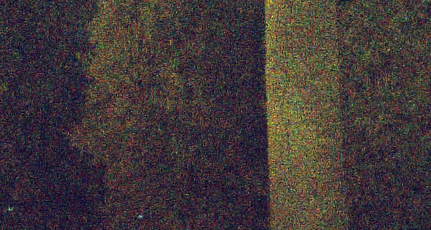}}  \hfill
        \subfloat{\includegraphics[width=0.249\textwidth]{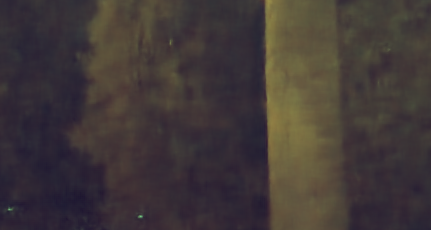}}  \hfill
        \subfloat{\includegraphics[width=0.249\textwidth]{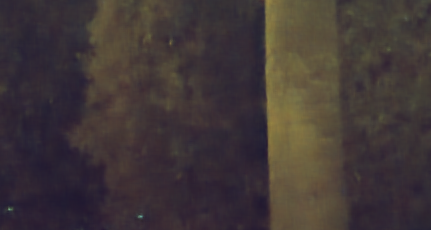}}  \hfill
        \subfloat{\includegraphics[width=0.249\textwidth]{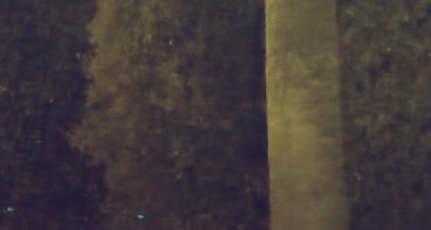}} \\
        \setcounter{subfigure}{0}
        \subfloat[real noisy]{\includegraphics[width=0.249\textwidth]{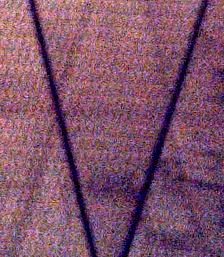}}  \hfill
        \subfloat[model-supervised]{\includegraphics[width=0.249\textwidth]{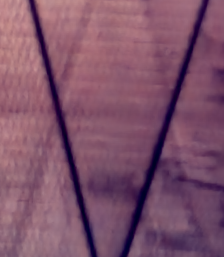}}  \hfill
        \subfloat[MF2F]{\includegraphics[width=0.249\textwidth]{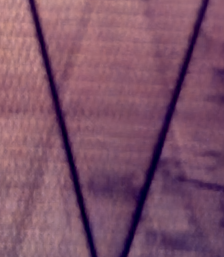}}  \hfill
        \subfloat[Self-supervised blind-spot]{\includegraphics[width=0.249\textwidth]{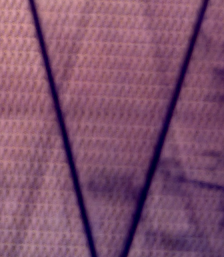}}
	\caption{Comparison of video denoising networks trained with supervision on synthetic data (b) or self-supervision on real data (c-d). All network architectures are based on UDVD~\cite{udvd}, MF2F (c) uses the self-supervised framework of~\cite{MF2F} and blind-spot (d) uses~\cite{udvd}. \emph{(top-brick wall ISO 3200)} Self-supervised networks recover more details. \emph{(middle top-bench ISO12800)} Natural texture of the stones and the granularity of the ground are removed by the supervised network. \emph{(middle bottom-trees ISO 3200)} Self-supervised networks have a better reconstruction of the texture of the trees \emph{(bottom-wire-grid ISO12800)} The structure of the wire grid is better reconstructed with the self-supervised networks.}
	\label{fig:main_results_expIII}
\end{figure*}

\section{Introduction}


For both images and videos, denoising is still an active research subject. All the more so in the case of real noise, where the real distribution of the noise may be unknown or at least hard to model.
In recent years, data-driven methods based on training convolutional neural networks (CNNs) have taken over the state of the art in several image and video restoration tasks. In addition to their superior performance, CNNs offer a greater flexibility as they can be trained to denoise potentially any type of noise~\cite{18-chen-see-in-the-dark,wang2017sar,kang2017deep,chang2019-blind-medical-denoising}. In contrast, traditional model-based approaches require a tractable model of the noise, and specific algorithms for each type of noise, \emph{e.g.}~\cite{lebrun2015noise,18-gonzales-denoising-decompression,maggioni2014joint,salmon2014poisson,coupe2009nonlocal,boulanger2009patch,zhao2019ratio}. This makes learning-based approaches an ideal candidate for restoration of real videos. Yet, this is still rather unexplored, with most of the research that considers real data focusing on single image denoising. The main reason for this is the lack of available training datasets for video denoising.

The standard approach {to data-driven methods is via} supervised {learning}, for which a dataset of pairs of input and expected output is used to train
the network. Network architectures trained in a supervised manner yield state-of-the-art results. However, supervised training requires large datasets with pairs of clean-noisy signals, which are very hard to obtain in the case of real images~\cite{18-chen-see-in-the-dark,plotz2017benchmarking,SIDD}, and even more so for real videos. The classical solution is to train networks on synthetic datasets where a clean video is artificially degraded. Nonetheless, CNNs are very sensitive to mismatches between the data distributions at training and testing times~\cite{plotz2017benchmarking}. Addressing this issue is currently one of the most important problems in learning-based image restoration, and has recently attracted a lot of attention.

One research trend focuses on generating realistic synthetic datasets for supervised training. 
The noise in the raw sensor is commonly approximated as an additive heteroscedastic Gaussian with a signal dependent variance~\cite{foi2008practical}.
A more accurate model is the Poisson-Gaussian model~\cite{foi2008practical}, which still has some limitations as it does not take into account non-linear behavior of the sensor (\emph{e.g.} clipping), dead pixels, heavy tails of read noise, \textit{etc}.
It has been shown that more comprehensive models of the noise yield better results~\cite{wei2021physics}. Other works rely on data driven generative approaches to synthesize noise~\cite{chen2018-GAN-noise-modelling,kim2019grdn, chang2020learning,yue2020dual,noiseflow, wolf2021deflow}.
Creating synthetic datasets requires synthesizing the clean data too. This is straightforward for RGB denoising, but it is far from trivial for raw denoising  \cite{brooks2019unprocessing,zamir2020cycleisp,wang2020practical} or for other imaging modalities.

Another research trend is based on developing self-supervised approaches that do not require any ground truth, \textit{i.e.} they can be trained directly on the degraded data. An additional advantage is that, in principle, no complex noise modeling is required in order to apply these methods.  The main principle of these techniques is to exploit the signal regularity, and train the network to predict one part of the signal from the rest. Self-supervised methods exist for both images~\cite{ehret2019model,batson19aN2S,noise2void,Moran2020noisier2noise,Quan2020self2self,dalsasso2021exploiting} and videos~\cite{MF2F,udvd} denoising, demosaicing~\cite{ehret2019join} and super-resolution~\cite{nguyen2021self,nguyen2022self}.
On artificial datasets, supervised training outperforms self-supervised approaches. However, recent self-supervised methods have shown competitive results, specially in video denoising.

The natural question is then, {\em what is the best approach to train denoising neural networks for real videos?}
Is it better to train with ground truth supervision on realistic synthetic datasets, or should one train directly on the real data with a self-supervised approach?
The former suffers from the generalization gap between simulated and real data, while the latter pays the price of not having supervision from a clean ground truth. The question  is which is the lesser evil. 


\paragraph{Contribution.} 
In this paper, we study the question of which training framework has to be used for video denoising networks: supervised on synthetic data or self-supervised on real data. 
This requires to compare quantitatively and fairly both approaches in a controlled setting. Ideally, this should be done by testing them on evaluation datasets of real natural videos with ground-truth. However, there are no such datasets due to the inherent complexity of simultaneously acquiring noisy and noiseless videos for natural dynamic scenes. 
We circumvent this problem by considering two surrogates for real data: 1) a synthetic raw dataset with a comprehensive noise model, and 2) a real dataset of static scenes for which ground truth can be estimated via frame averaging. 
Finally, we evaluate both approaches on real natural videos visually. In all cases, we apply a rigorous training methodology to make sure that we compare fairly the training approaches. 

The next section reviews the related work. In Section~\ref{sec:self-supervised_training_methods}, we present the architecture used in this study as well as a description of the self-supervised trainings. The overall protocol of the study (including datasets and training strategies) is detailed in Section~\ref{sec:methodology}. Experiments details and results are presented in Section~\ref{sec:experiments}.

\section{Related work}


Self-supervised training methods are often compared to supervised training on synthetic datasets \cite{ehret2019model,batson19aN2S,noise2void,MF2F,udvd}. In this setting, supervised training is optimal (e.g. with respect to the MSE) and the goal is to achieve the same performance with self-supervision. Our situation is different, since we are interested on the performance on real data of a supervised network trained on synthetic data.

In the case of still images, it is possible to acquire real datasets with ground truth. The ground truth can be either estimated by acquiring a burst of images of static scenes and averaging them \cite{SIDD,18-chen-see-in-the-dark} or using long exposure times \cite{plotz2017benchmarking}. 
In such datasets it is possible to train with supervision on real data, and it has been observed that training a network with unrealistic simulated data leads to worse results~\cite{noiseflow,wei2021physics}.




This motivated research into how to better simulate real noise. 
The simplest noise model for raw images is the heteroscedastic Gaussian noise model \cite{liu2014practical} which supposes the noise to be additive, zero-mean and with a variance as a affine function of the intensity. This corresponds to the sum of two noise sources: the shot noise modeling photons arriving at the sensor and the readout noise introduced by the electronics. In spite of its known limitations, this model is widely used~\cite{guo2018toward,plotz2018neural,jaroensri2019generating,brooks2019unprocessing,zamir2020cycleisp,wang2020practical}.
In \cite{claus2019videnn} the authors use an additive and zero-mean heteroscedastic Gaussian noise but the variance does not follow an affine model. In~\cite{zhu2016noise,zhao2014robust} a Gaussian mixture model is used. 
In \cite{wei2021physics}, the shot noise is considered Poissonian and a Tukey-Lambda distribution is used to model heavy tails in the readout noise. Additionally, other noise sources are also modeled like the banding pattern noise (e.g. row noise) or quantization noise. 

Other approaches for simulating real noisy sequences use data-driven generative methods, such as adversarial generative models~\cite{chen2018-GAN-noise-modelling,kim2019grdn,yue2020dual}. 
In these works, a generative network is trained to generate a noise close to the real one while a discriminative network is trained to determine whether a noise sample is real or has been generated. 
In \cite{noiseflow} a neural network entirely composed of invertible layers is used to simulate realistic noise from clean data. It was trained on the SIDD dataset~\cite{SIDD} and can reproduce the realistic noise of the five cameras with a smaller KL-divergence with respect to the real noise than the heteroscedastic Gaussian noise. Similarly in~\cite{wolf2021deflow}, a CNN is trained to generate realistic degraded data from clean ones.

For raw denoising, it is important also to simulate the raw clean ground truth, a problem that has received less attention. The standard approach is to use sRGB images and apply a simple inverse camera pipeline to generate the raw~\cite{guo2018toward,brooks2019unprocessing}. In~\cite{zamir2020cycleisp} the inverse pipeline is implemented by a network that is learned from real data.

\section{Self-supervised training methods}\label{sec:self-supervised_training_methods}


Self-supervised learning methods learn directly from the noisy data by exploiting differences in the correlation structure of signal and noise. A part of the input is withheld from the network, which is trained to predict the withheld part. If the noise of the withheld part is independent from the one given to the network, then the network can only minimize the loss by predicting the clean signal.



The state of the art in self-supervised video denoising is represented by \emph{Multi Frame-to-Frame} (MF2F)~\cite{MF2F} and \emph{Unsupervised Deep Video Denoising} (UDVD)~\cite{udvd}.
Both techniques were found to be efficient on the real raw video denoising task, and thus we are going to include both of them in our comparison. We will describe them below.

\subsection{UDVD: blind-spot network for video denoising}
The UDVD method relies on the blind-spot technique recently introduced in~\cite{noise2void,batson19aN2S,BSS,krull2019probabilisticN2V}: 
a special convolutional network architecture is used, which has a blind-spot at the center of its receptive field. The network is trained to predict the value of this pixel in the noisy video. It is predicted from the surrounding neighbors (both spatial and temporal), exploiting the spatio-temporal regularity of the clean video. The blind-spot technique decreases the denoising performance compared with a non blind-spot (normal) network, as many details are lost. This gap is significant in the case of image denoising, but it is less discernible for video. 

\begin{figure}
    \centering
    \includegraphics[width=\linewidth]{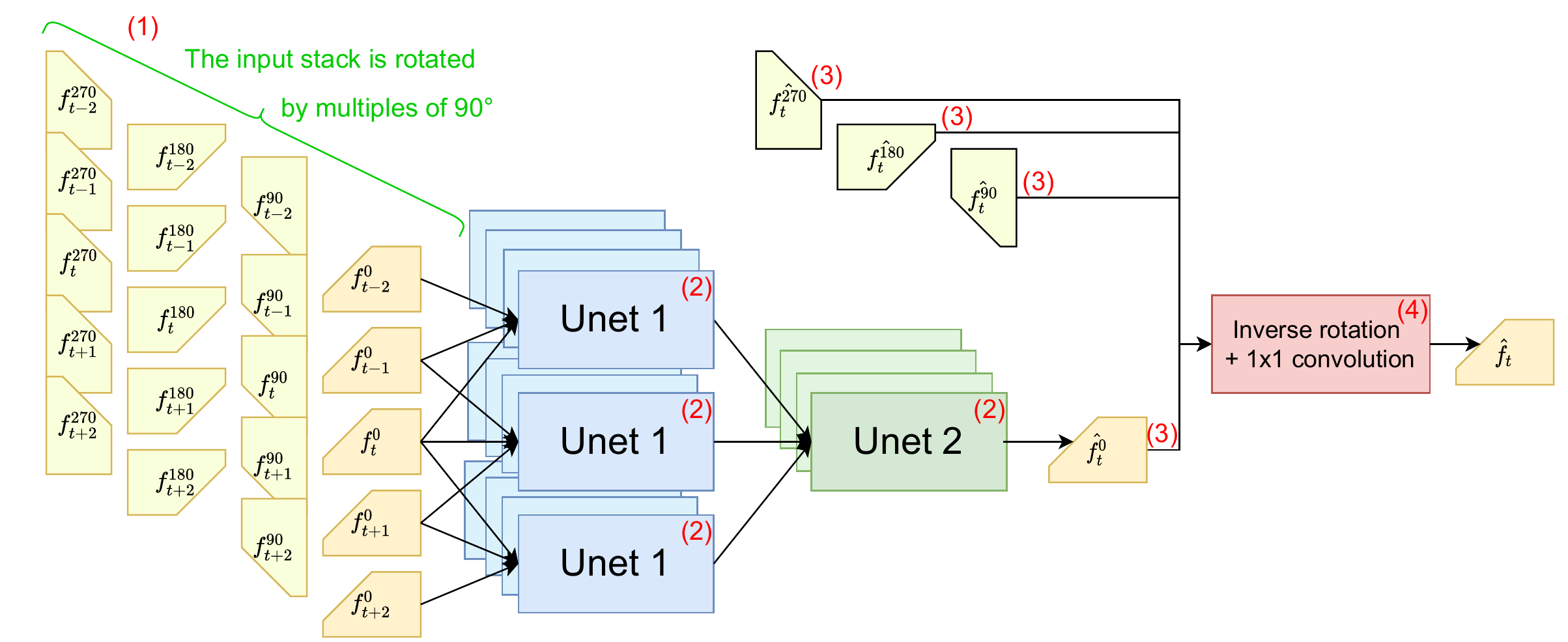}
    \caption{Architecture of the network introduced in~\cite{udvd}. The input stack is firstly rotated by multiples of 90° (1). Each four rotated stacks is processed by the cascaded Unets (2), producing four outputs (3) which are combined together after the rotations are inverted and a $1\times1$ convolution (4).}
    \label{fig:UDVD_architecture}
\end{figure}

\paragraph{Network architecture.} For our experiments, we use the architecture of UDVD~\cite{udvd}. This video denoising network takes a stack of five frames as input. It consists of two cascaded Unets (as in \cite{tassano2019fastdvdnet}). The first Unet is applied three times on each group of three contiguous frames. This produces three outputs that are fused by a second Unet into the final denoised output. The input stack is rotated by the four multiples of 90° and denoised by the network. The four outputs are finally combined by a $1\times1$ convolution. The architecture is shown in Figure~\ref{fig:UDVD_architecture}. 

To generate the blind-spot, the first Unet uses asymmetric convolutional filters that are vertically causal so that the four outputs only depend on the pixels above. In this way, the receptive field does not contain the central pixel. The blind-spot can be removed by shifting the input data one pixel down after the rotation. The UDVD architecture is also bias-free~\cite{mohan2019robust}, which has proved to generalize better to unseen noise level at test time.

\paragraph{UDVD training.}

The self-supervised UDVD blind-spot network is trained by minimizing the L2 loss between the output of the network and the corresponding noisy input frame. 

\subsection{MF2F training}\label{sec:train_MF2F}

In MF2F, the weights $\theta$ of a network $\mathcal{F}_\theta$ are updated by minimizing the loss $\Vert \kappa_t \circ \left( W_{t, t-1} \mathcal{F}_\theta \left( \mathcal{S}_t\right) - f_{t-1} \right) \Vert_1$, where $\kappa_t$ is an occlusion mask, $W_{t, t-1}$ is a warping operator from frame at time $t$ to $t-1$ (based on an estimated optical flow), $\mathcal{S}_t$ is the stack of frames $[f_{t-4}, f_{t-2}, f_t, f_{t-2}, f_{t+4}]$ and $f_{t-1}$ is the first past frame serving as target (to prevent the trivial identity mapping, it is out of the input stack). The alignment requires a high quality optical flow plus a mask for alignment errors, which are estimated on the noisy data. The MF2F results strongly depends on the optical flow accuracy, which is computed using the TV-L1 method~\cite{zach2007duality,perez2013tv}. 

The application of the warping operator $W_{t,t-1}$ requires interpolating the network output at subpixel positions. Interpolating the raw image is problematic. A naive approach would be to pack the raw as a 4 channels image of half the resolution and warp each channel. However, these low resolution channels are heavily aliased. We found better results applying a demosaicing $D$ to the network output, warping on the RGB domain, and re-mosaicing it afterwards. That is, our warping operator can be expressed as
$W^{\text{raw}}_{t,t-1} = MW^{\text{rgb}}_{t,t-1}D$, where $M$ is the remosaicing operator. For the demosaicing we use the Hamilton-Adams method~\cite{jin2021mathematical,Hamilton1997}.

\section{Methodology}\label{sec:methodology}

\begin{figure*}[t]
    \centering
    \subfloat[Pre-training on the \emph{synthetic dataset}]{\includegraphics[width=0.5\textwidth]{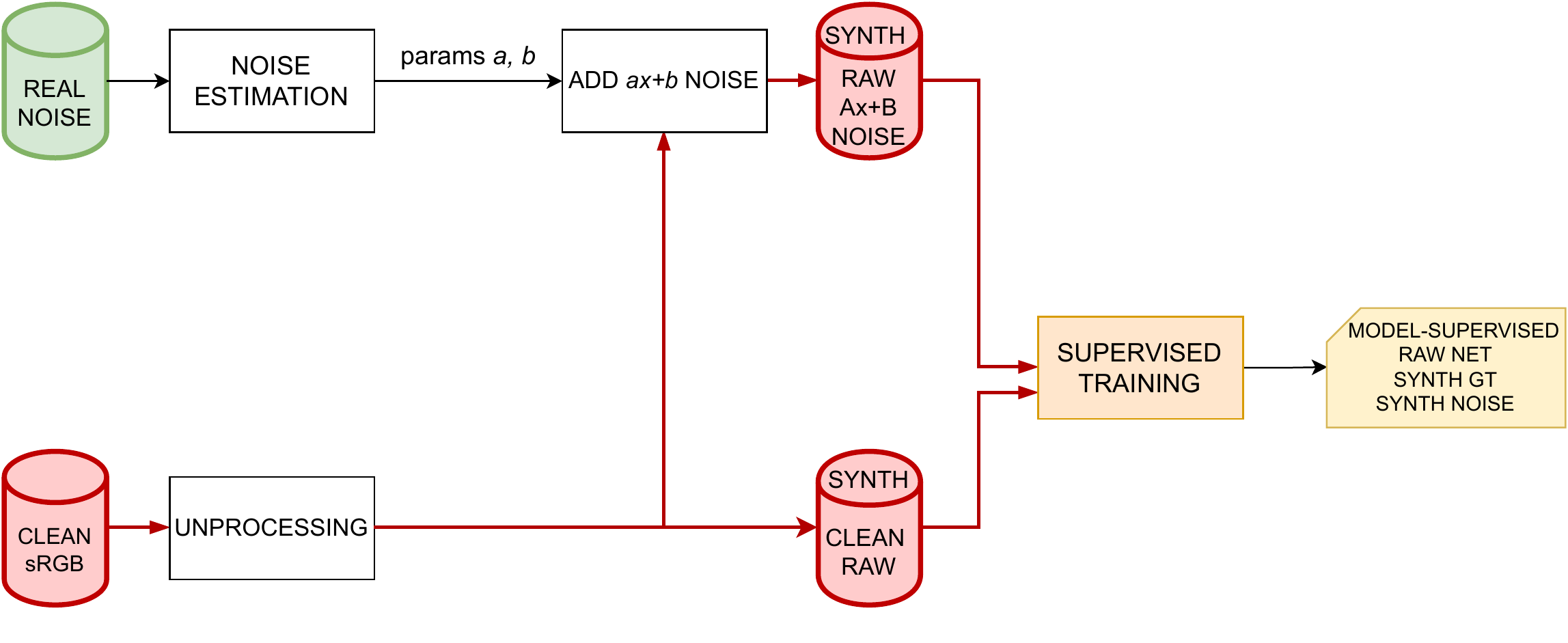}}%
    \subfloat[Fine-tuning on the \emph{surrogate dataset}]{\includegraphics[width=0.5\textwidth]{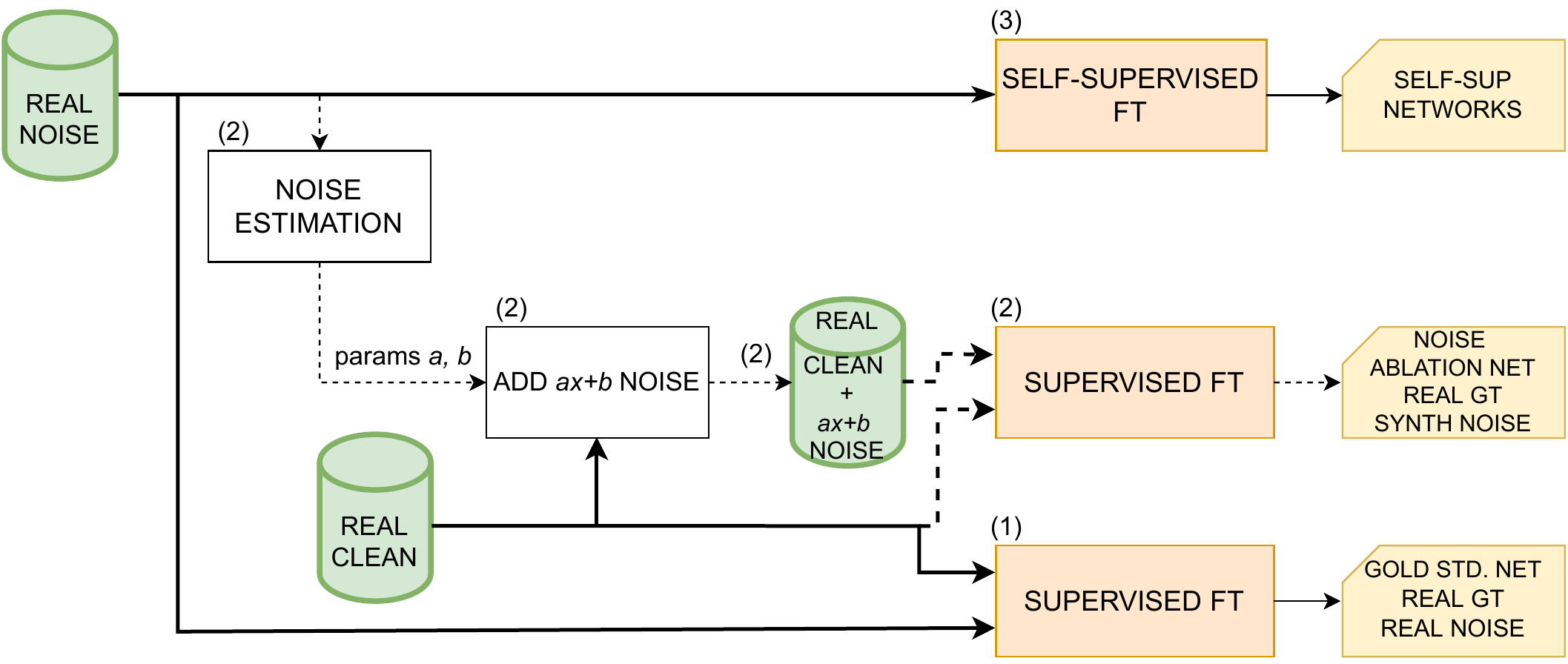}}
    \caption{The \emph{surrogate real dataset} is the \textcolor{green}{green} cylinder, the \emph{synthetic dataset} is in \textcolor{red}{red}. (a) The model-supervised is trained with supervision on the \emph{synthetic dataset} with synthetic noise (either with or without blind-spot). (b) The previous model-supervised are fine-tuned on the \emph{surrogate dataset}. The steps are (1) fine-tune on real data (2) (when possible) fine-tune on real clean data but with synthetic noise (3) Self-supervised fine-tuning directly on noisy data (UDVD and MF2F).}
    \label{fig:diagram_methodology}
\end{figure*}

Our goal is to compare two strategies for training a denoising network for raw real videos: supervised training on realistic synthetic data, or self-supervised training directly on the real data. To that aim we need a dataset of synthetic noisy videos and one of real natural videos for evaluation. In the following, we describe our evaluation protocol (see Figure \ref{fig:diagram_methodology}).

\paragraph{Dataset of real videos.}
Since there are no datasets of real natural videos with ground truth, we will consider two surrogates: (1) synthetic videos with a comprehensive noise model, and (2) static real videos with ground truth generated by frame averaging. The first will allow us to measure the effect of an oversimplified noise model in the \emph{synthetic dataset} of dynamic scenes with natural motion. The second is static, but will be useful to have a quantitative evaluation on real data.
Lastly, we will consider a dataset of real natural videos for visual evaluation. More details about these datasets will be given respectively in Sections \ref{sec:expI}, \ref{sec:expII} and \ref{sec:expIII}.
For simplicity we will talk about the \emph{surrogate real dataset} (abridged to \emph{surrogate dataset}) in the following, even though it might not be actually real data, but our proxy for real data. The \emph{surrogate real dataset} is represented in green in the diagrams of Figure~\ref{fig:diagram_methodology}.

\paragraph{Dataset of synthetic videos.}
For each \emph{surrogate dataset} we generate a synthetic realistic dataset with clean-noisy pairs for supervised training (represented in red in Figure~\ref{fig:diagram_methodology}).
We use the REDS 120 dataset~\cite{reds} which consists of 270 dynamic sequences (split in 240 training and 30 validation sequences) of outdoors scenes taken in daylight conditions, with frame rate 120 fps and of size $1280 \times 720$. We temporally downsampled each sequence by taking one frame over three, resulting sequences with 166 frames at 40 fps.
Note that this makes the task more complicated for MF2F whose results highly depend on the optical flow estimation accuracy, and therefore on the amount of motion.
    
These clean sRGB sequences are unprocessed back to the raw domain following~\cite{brooks2019unprocessing}, which we adapted to our specific case. First, we use a fix color correction matrix throughout all the dataset. We sample random white balance coefficients for each sequence, and use the same coefficients for all frames in the sequence. In the supplementary material, we give more details about our unprocessing procedure steps.
This gives us a dataset of clean raw video sequences.

Finally, we add realistic noise to the unprocessed ground-truth for simulating real noisy sequences from the clean ones. For that purpose, a heteroscedastic Gaussian noise model is estimated from the \emph{surrogate dataset}~\cite{liu2014practical}. More details about the noise estimation can be found in the supplementary material.

Note that the \emph{synthetic dataset} is tailored to model the \emph{surrogate dataset}: we use the same Bayer pattern, the ranges of both datasets are matched and the parameters of the synthetic noise model are fitted to approximate the noise in the \emph{surrogate dataset}.

\subsection{Networks}

We will use for our experiments the UDVD architecture described in the previous section. This network is computationally costly and has a significant memory footprint. In this paper, we do not focus on achieving the state of the art and reduce this architecture by a factor 4 by using 1/4 of the channels in all layers. 
This architecture can be used with or without the blind-spot. 

Due to the small size of the \emph{surrogate dataset}, we followed~\cite{yue2020supervised,MF2F} and pre-trained the network with supervision on the bigger \emph{synthetic dataset}.

We pre-trained this architecture with a blind-spot as well as without the blind-spot (denoted as \emph{normal}). The reason is that we do not need a blind-spot network for applying MF2F as well as for other supervised training strategies discussed later; while the self-supervised UDVD requires a blind-spot.


For comparing the supervised and self-supervised frameworks, we consider different training strategies. Figure~\ref{fig:diagram_methodology} summarizes them. 
Note that once trained, the evaluation of networks trained with or without supervision requires the same amount of time and computational resources.
We now describe the different networks and how we trained them.

\paragraph{Model-supervised net} This network is trained with supervision on the \emph{synthetic dataset}. We train two versions of this network: normal and with the blind-spot.
The latter will be used as the pre-trained network for the self-supervised blind-spot fine-tuning, 
while the former is \emph{the supervised network trained on synthetic data that we wish to compare with the self-supervised approaches.} 

\paragraph{Gold standard net} The \emph{gold standard} solution for such training is to train with supervision directly on the real data. Although this is not possible in practice because it requires to have access to a large dataset with clean-noisy pairs, it is possible here to fine-tune the \emph{normal model-supervised net} on the \emph{surrogate dataset}. This will give us a reference of the best training that could be achieved to situate the performance of the other trainings.

\begin{figure*}[h!]
    \centering
        \subfloat{\includegraphics[width=0.165\textwidth]{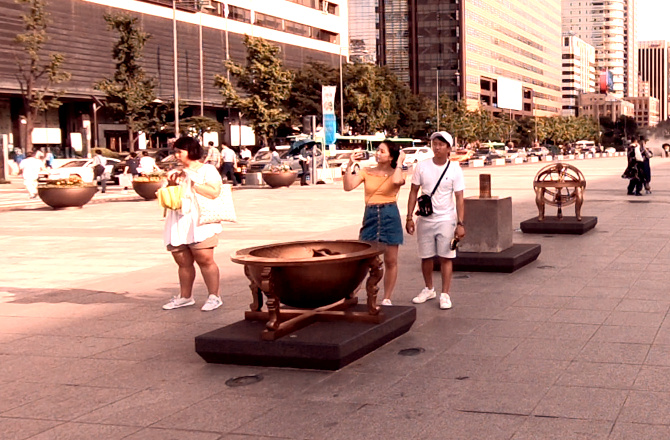}}  \hfill
        \subfloat{\includegraphics[width=0.165\textwidth]{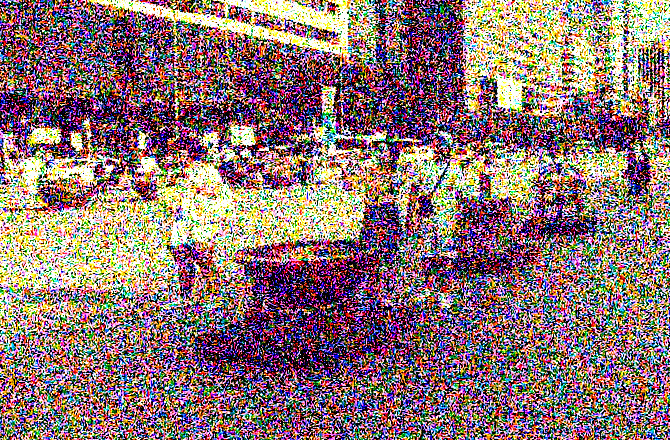}}  \hfill
        \subfloat{\includegraphics[width=.165\textwidth]{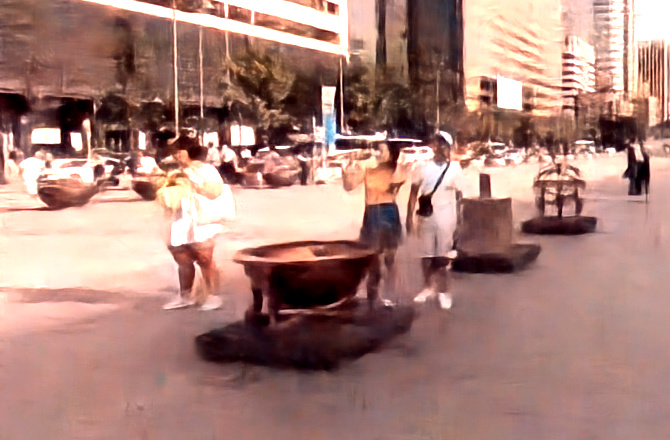}}  \hfill
        \subfloat{\includegraphics[width=.165\textwidth]{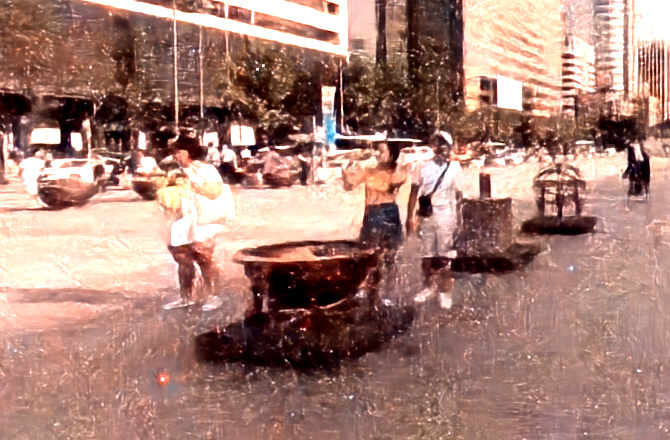}}  \hfill
        \subfloat{\includegraphics[width=0.165\textwidth]{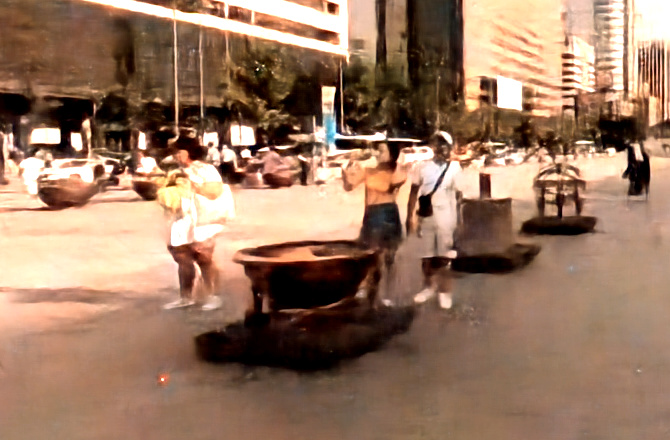}}  \hfill
        \subfloat{\includegraphics[width=.165\textwidth]{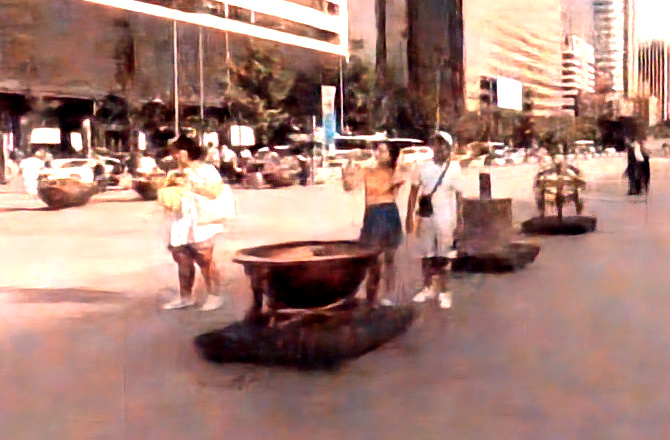}}\\
        \subfloat{\includegraphics[width=0.165\textwidth]{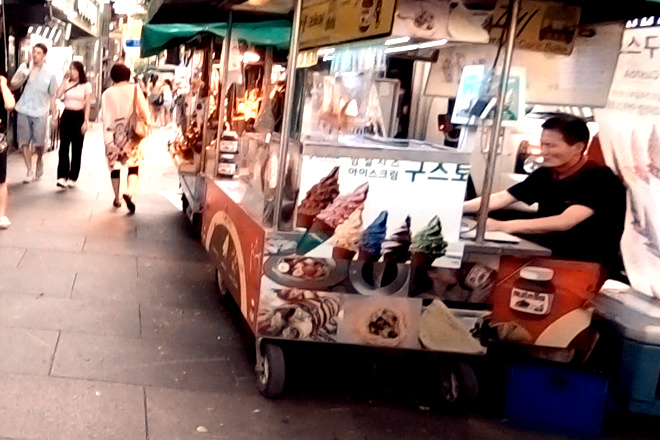}}  \hfill
        \subfloat{\includegraphics[width=0.165\textwidth]{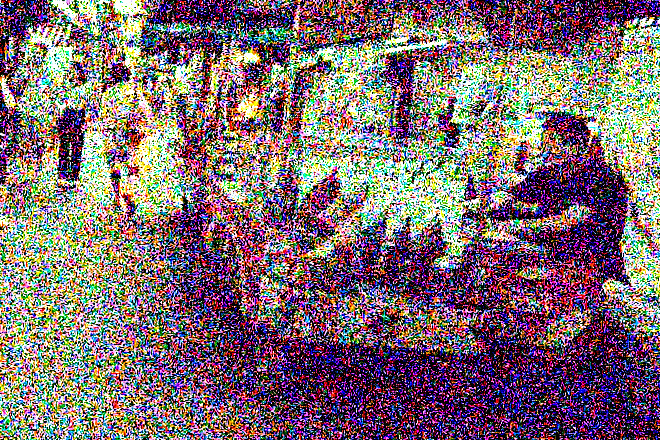}}  \hfill
        \subfloat{\includegraphics[width=.165\textwidth]{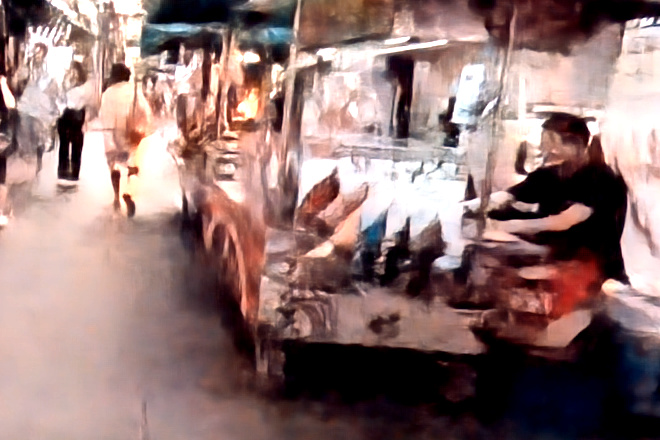}}  \hfill
        \subfloat{\includegraphics[width=.165\textwidth]{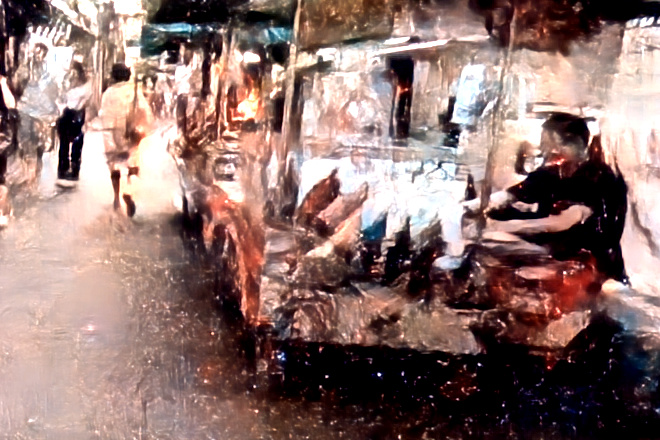}}  \hfill
        \subfloat{\includegraphics[width=0.165\textwidth]{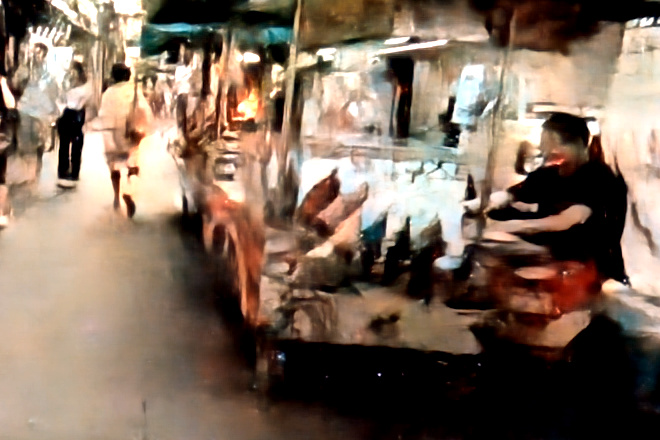}}  \hfill
        \subfloat{\includegraphics[width=.165\textwidth]{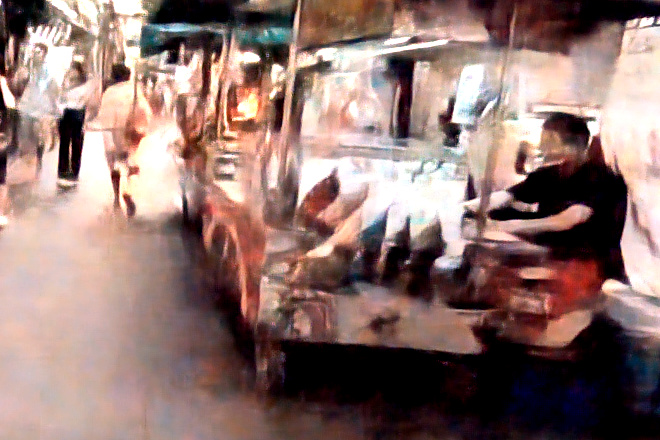}}\\
        \subfloat{\includegraphics[width=0.165\textwidth]{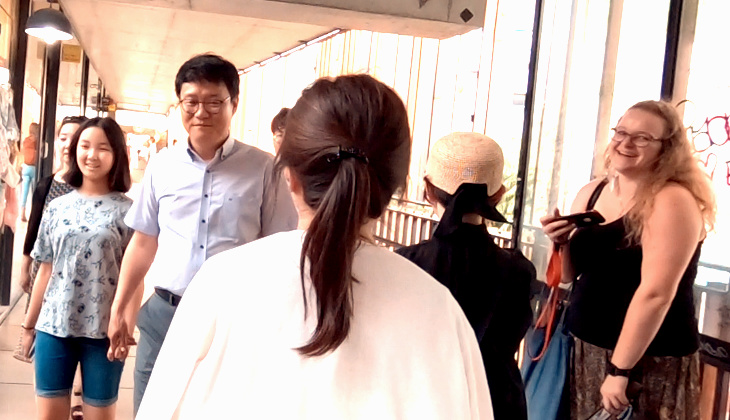}}  \hfill
        \subfloat{\includegraphics[width=0.165\textwidth]{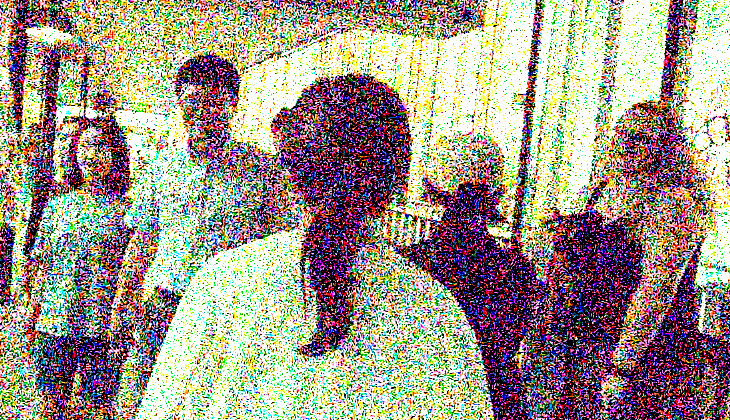}}  \hfill
         \subfloat{\includegraphics[width=.165\textwidth,]{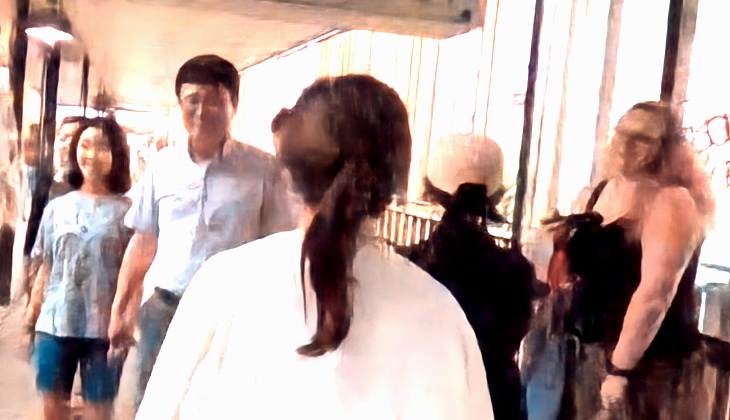}}  \hfill
        \subfloat{\includegraphics[width=.165\textwidth]{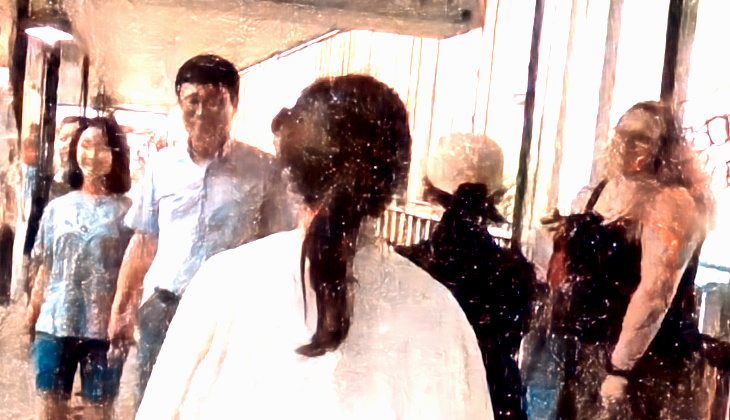}}  \hfill
        \subfloat{\includegraphics[width=0.165\textwidth]{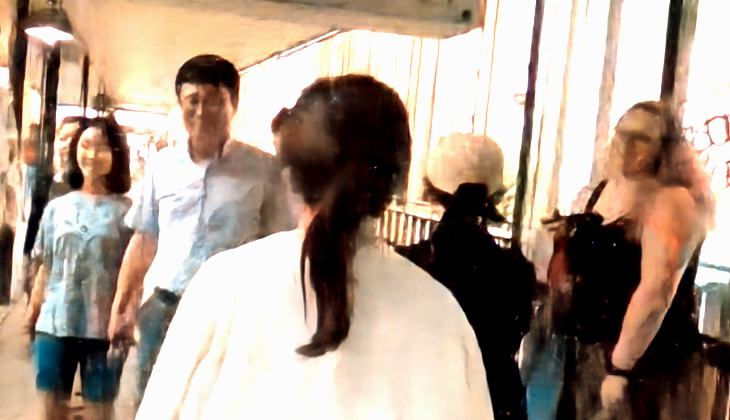}}  \hfill
        \subfloat{\includegraphics[width=.165\textwidth]{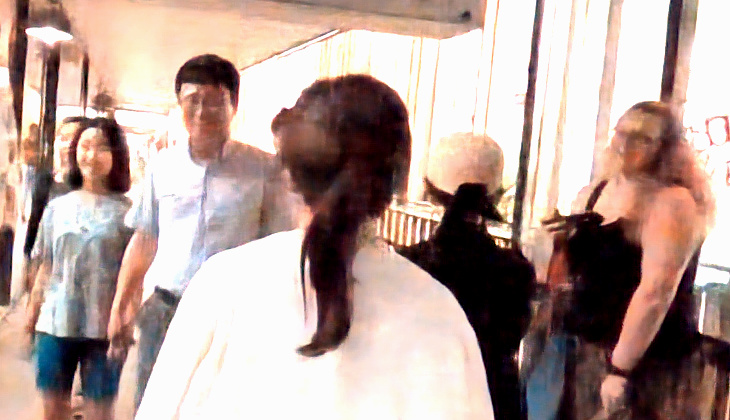}}\\
        \setcounter{subfigure}{0}
        \subfloat[gt]{\includegraphics[width=0.165\textwidth]{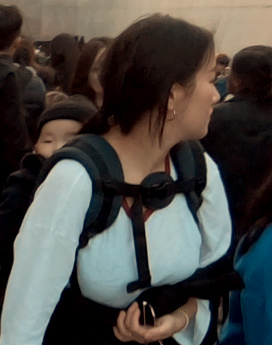}}  \hfill
        \subfloat[real noisy]{\includegraphics[width=0.165\textwidth]{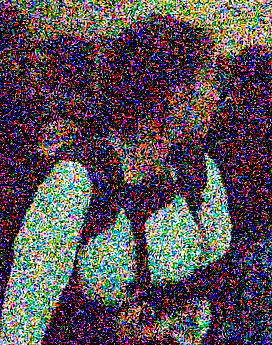}}  \hfill
         \subfloat[Gold standard]{\includegraphics[width=.165\textwidth,]{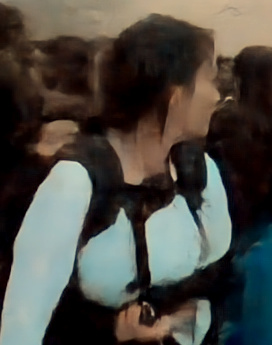}}  \hfill
        \subfloat[Model-sup. normal]{\includegraphics[width=.165\textwidth]{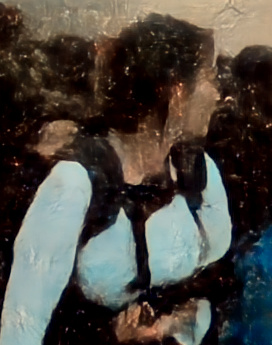}}  \hfill
        \subfloat[MF2F]{\includegraphics[width=0.166\textwidth]{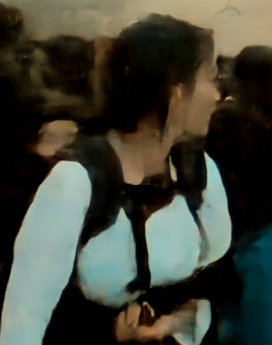}}  \hfill
        \subfloat[Self-sup. blind-spot]{\includegraphics[width=.165\textwidth]{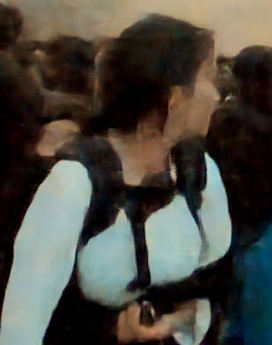}}\\
	\caption{Comparison of the different training strategies for the Exp I.}
	\label{fig:main_results_expI}
\end{figure*}

\paragraph{Noise-ablation net} Two kinds of modeling were used for the supervised trainings: the unprocessing of sRGB to generate synthetic raw clean videos and the noise. When possible, we fine-tune an \emph{noise-ablation net} that will allow us to differentiate the impact of the noise model from that of the generation of the clean data by eliminating the unprocessing step.
To this aim, we generate noisy images by adding synthetic heteroscedastic Gaussian noise to the clean ground truth of the \emph{surrogate dataset}. We fine-tune the normal model-supervised net on this data with supervision from the real ground truth.

\paragraph{Self-Supervised blind-spot} We fine-tune the pre-trained UDVD architecture with blind-spot on the \emph{surrogate dataset} with self-supervision following~\cite{udvd}.

\paragraph{MF2F net} A second self-supervised network is trained following the MF2F framework as explained in the Section~\ref{sec:train_MF2F}. Given that MF2F does not requires the network to \linebreak have a blind-spot, we use the weights of the pre-trained \emph{normal} \emph{model-supervised net} as starting point of the fine-tuning.

\section{Experiments}\label{sec:experiments}

\begin{table*}[t]
\centering
\setlength{\tabcolsep}{0.68em}
\begin{tabular}{ |c|c|c|c|c||c|c| }
\hline
  & \multicolumn{4}{c||}{Supervised networks} &  \multicolumn{2}{c|}{Self-supervised networks}\\
 \hline
  \multirow{2}{*}{Exp} & Gold standard & Noise-ablation & \multicolumn{2}{c||}{Model-supervised} & \multirow{2}{*}{MF2F} & \multirow{2}{*}{blind-spot}\\
   & net & net & normal & blind-spot  & & \\
   \hline
   I &  29.90 / .8393 & N/A & 28.77 / .7886 & 27.89 / .8862 & 28.76 / .8153 & 29.46 / .9325 \\
   II & 50.03 / .9913 & 50.03 / .9913 & 47.22 / .9847 & 46.55 / .9938 & 47.42 / .9849 & 48.71 / .9965 \\
   \hline 
\end{tabular}
\caption{Average PSNR and SSIM over all the sequences of the \emph{surrogate real} test dataset. The model-supervised normal do not have a blind-spot. The model-supervised with blind-spot serves as a pre-trained for the self-supervised UDVD. In Experience I, the \emph{surrogate dataset} is synthetic. Thus we cannot derive an noise-ablation net for this experiment as it would be identical to the model-supervised normal.}
\label{tab:main_results}
\end{table*}

\begin{figure*}
    \centering
        \subfloat[gt]{\includegraphics[width=0.249\textwidth]{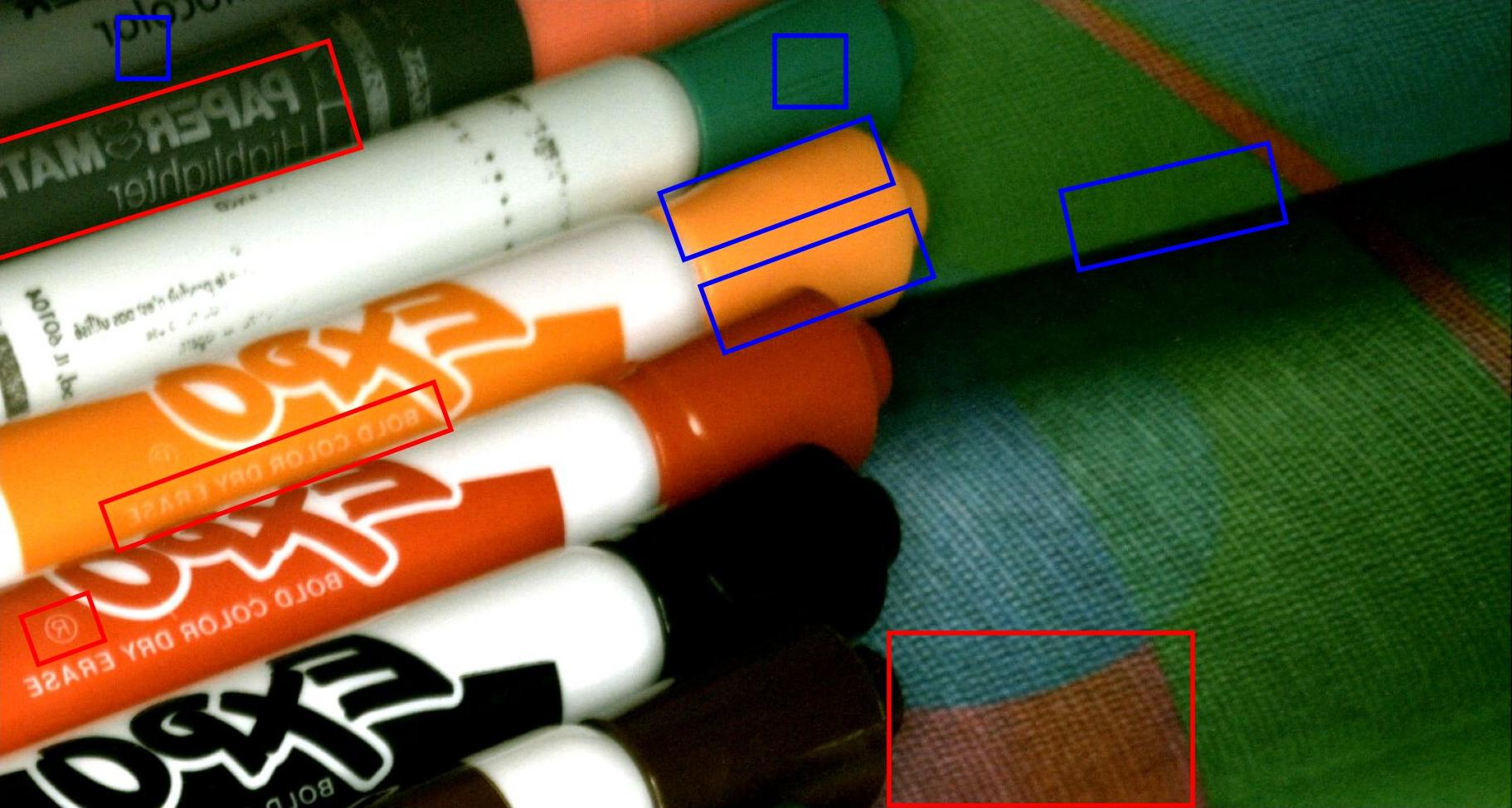}} \hfill
        \subfloat[real noisy]{\includegraphics[width=0.249\textwidth]{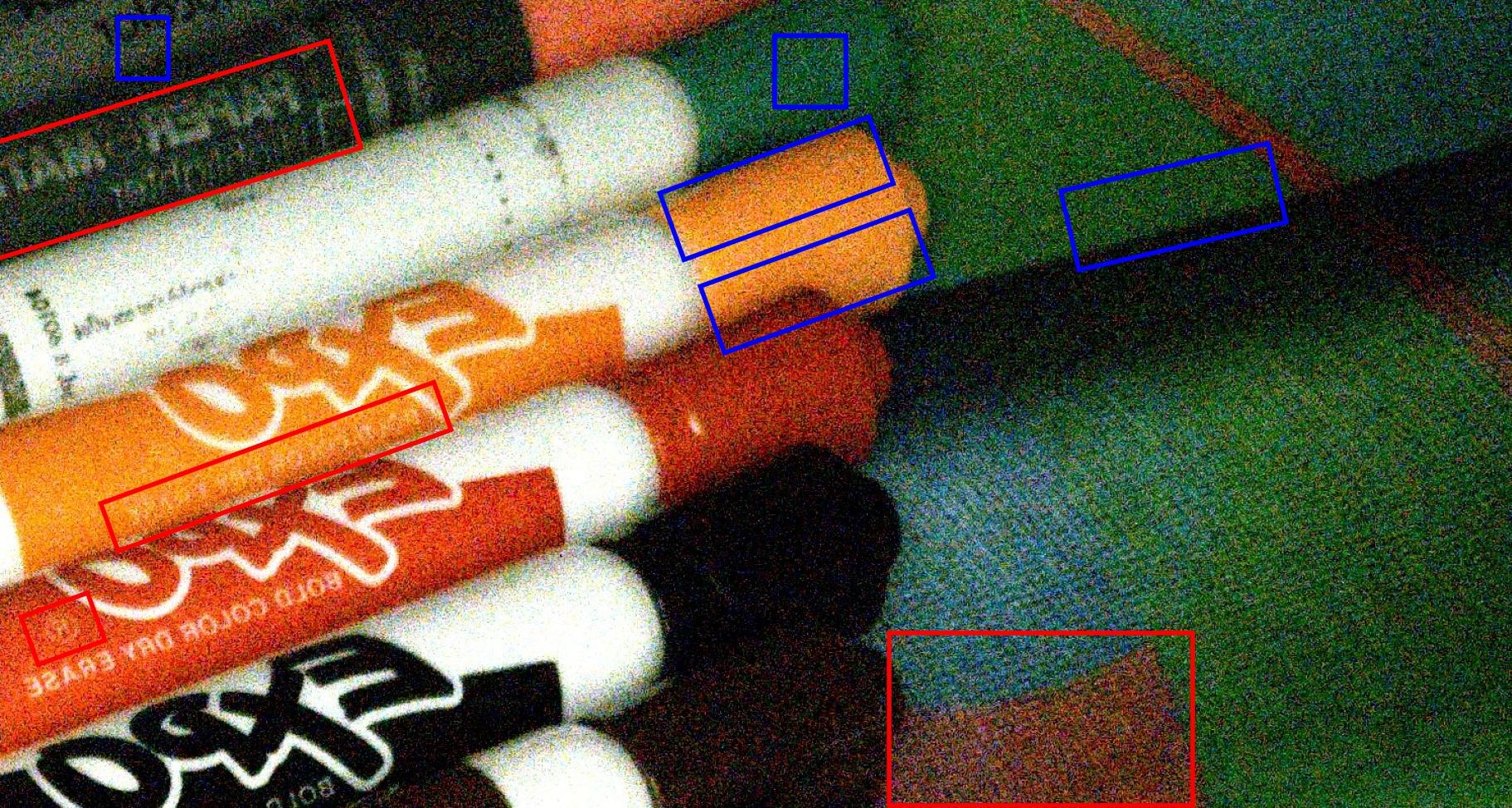}} \hfill
        \subfloat[Gold standard]{\includegraphics[width=0.249\textwidth]{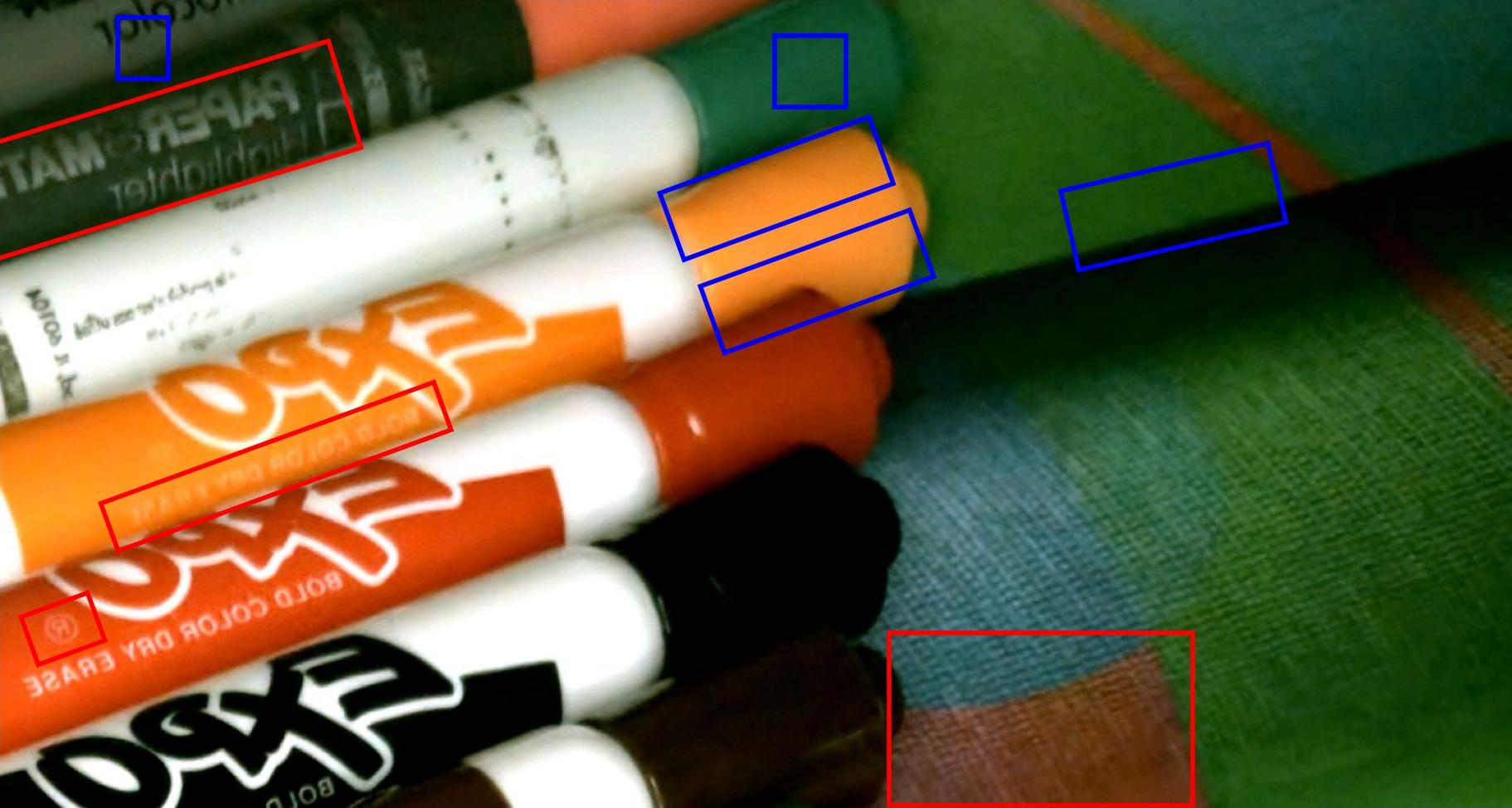}} \hfill
        \subfloat[Noise-ablation]{\includegraphics[width=.249\textwidth]{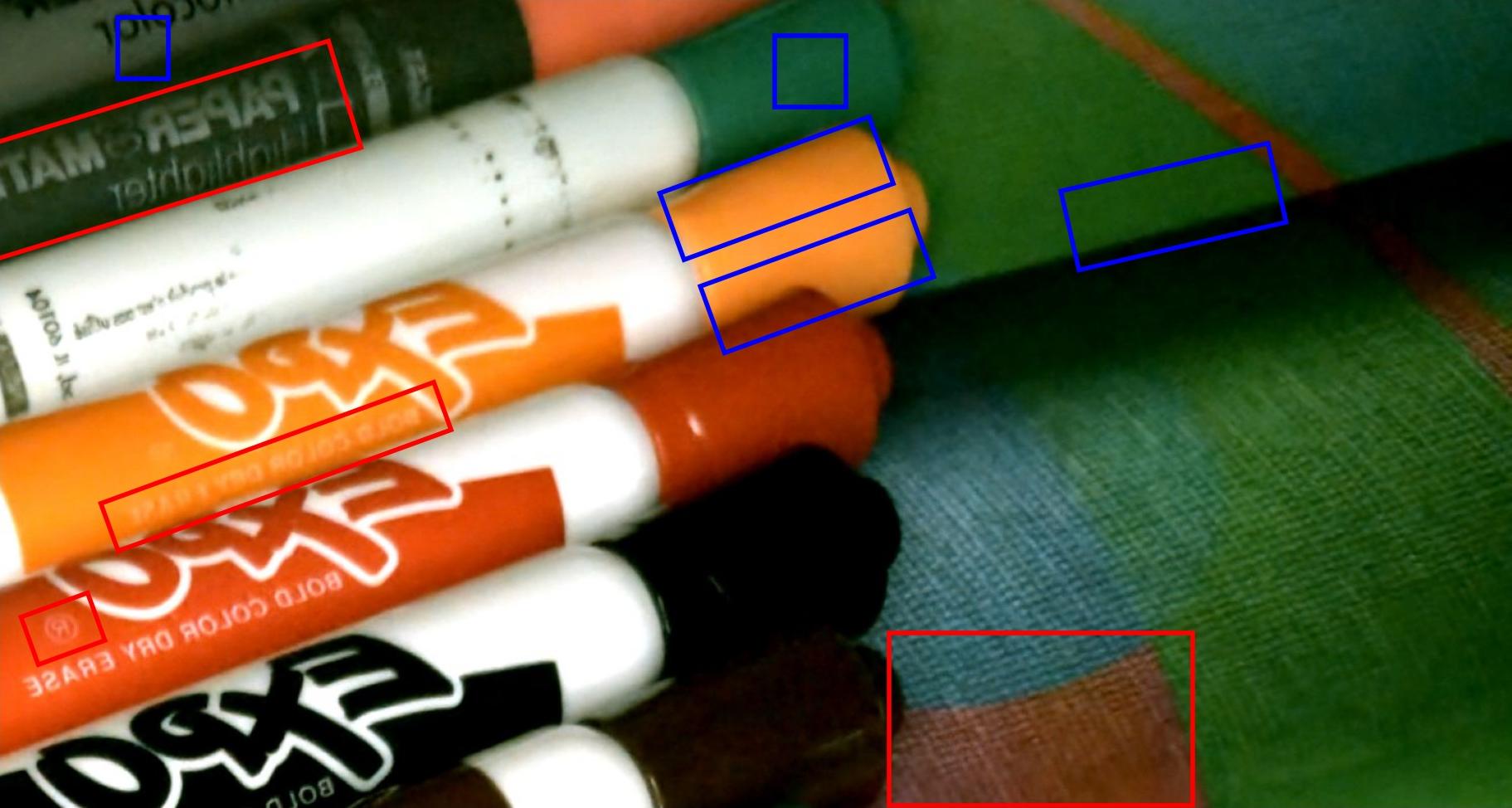}}\\
        \subfloat[Model-supervised normal]{\includegraphics[width=.249\textwidth]{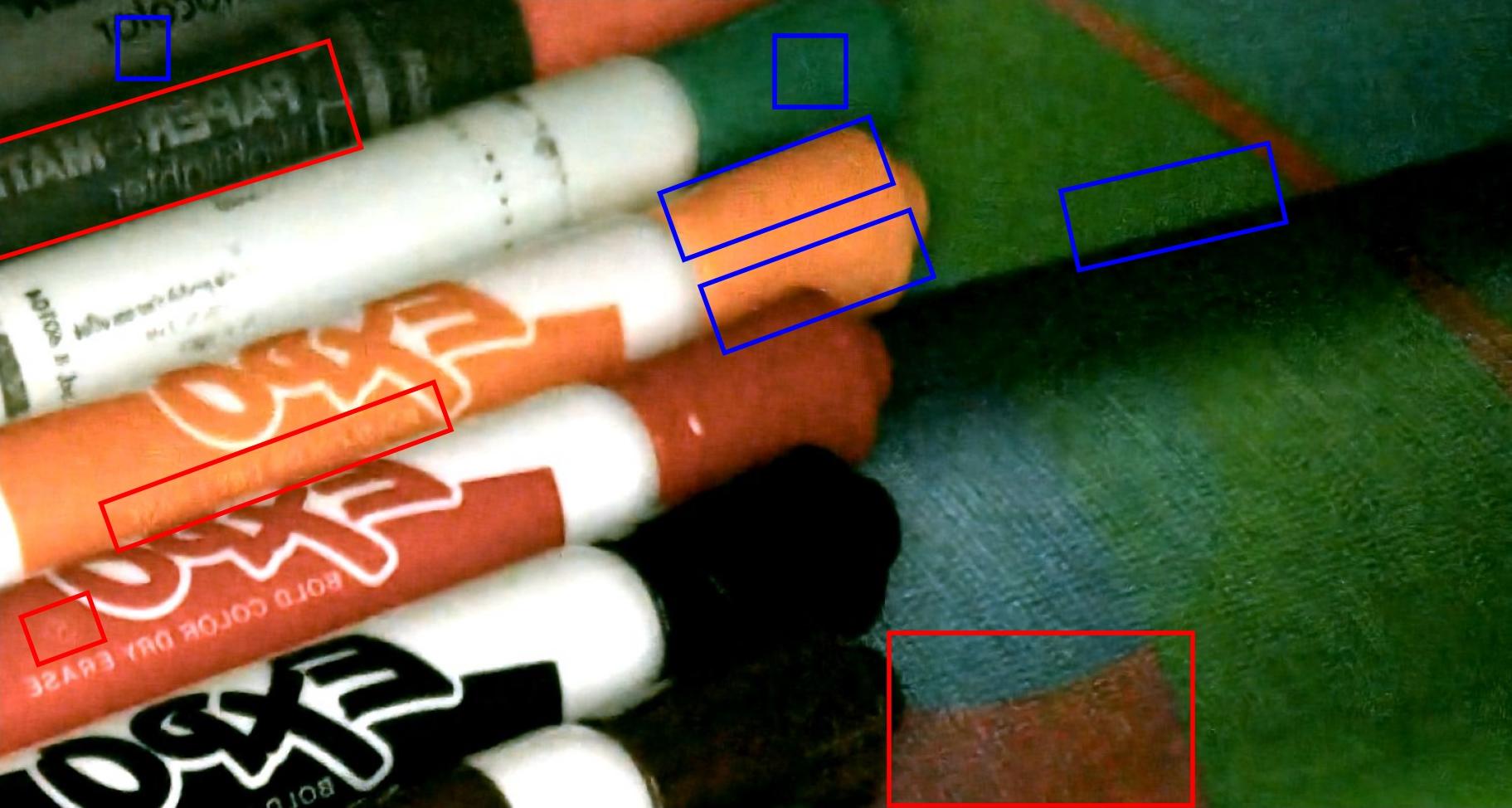}} \hfill
        \subfloat[Model-supervised blind-spot]{\includegraphics[width=.249\textwidth]{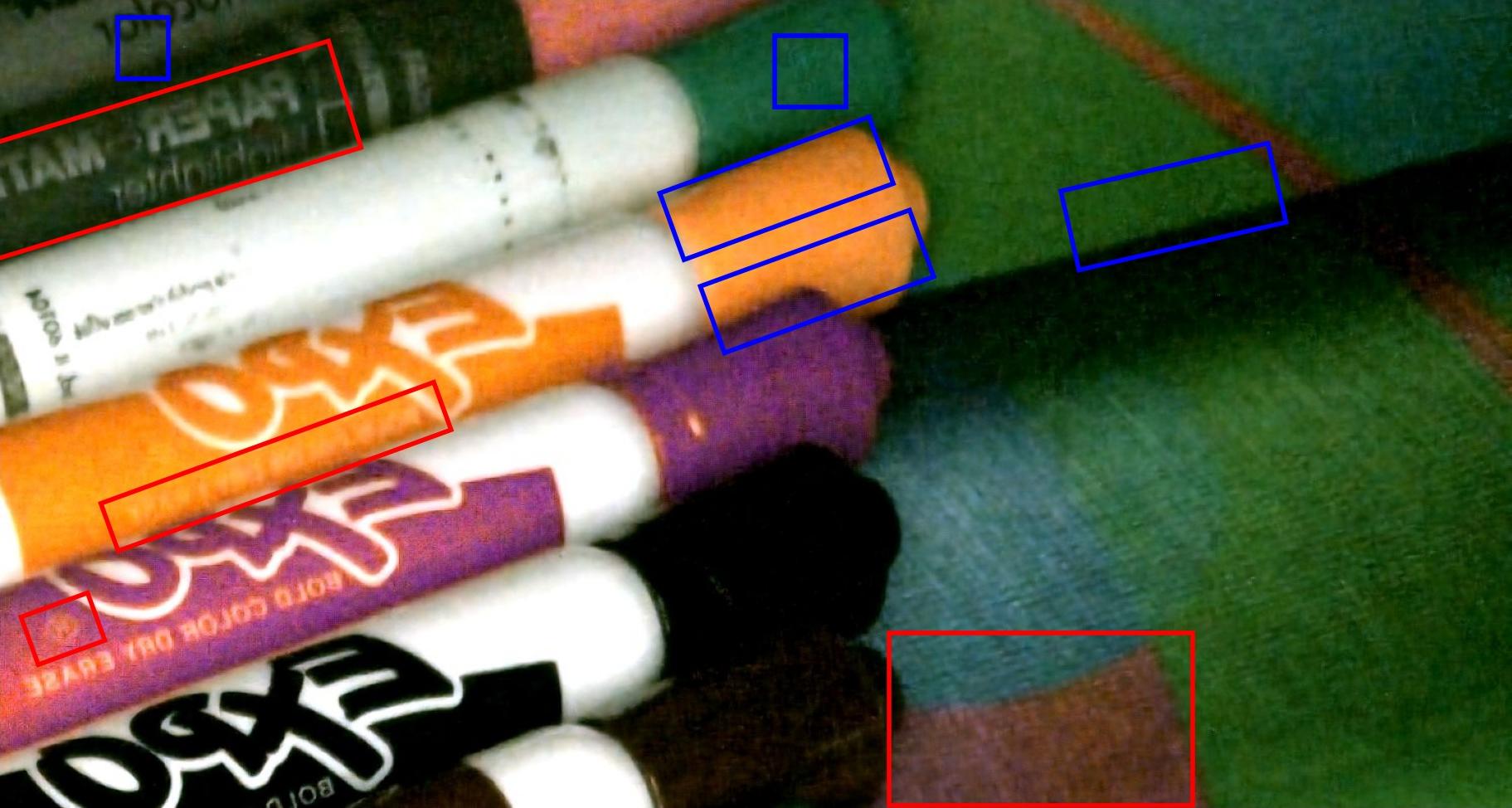}} \hfill
        \subfloat[MF2F]{\includegraphics[width=0.249\textwidth]{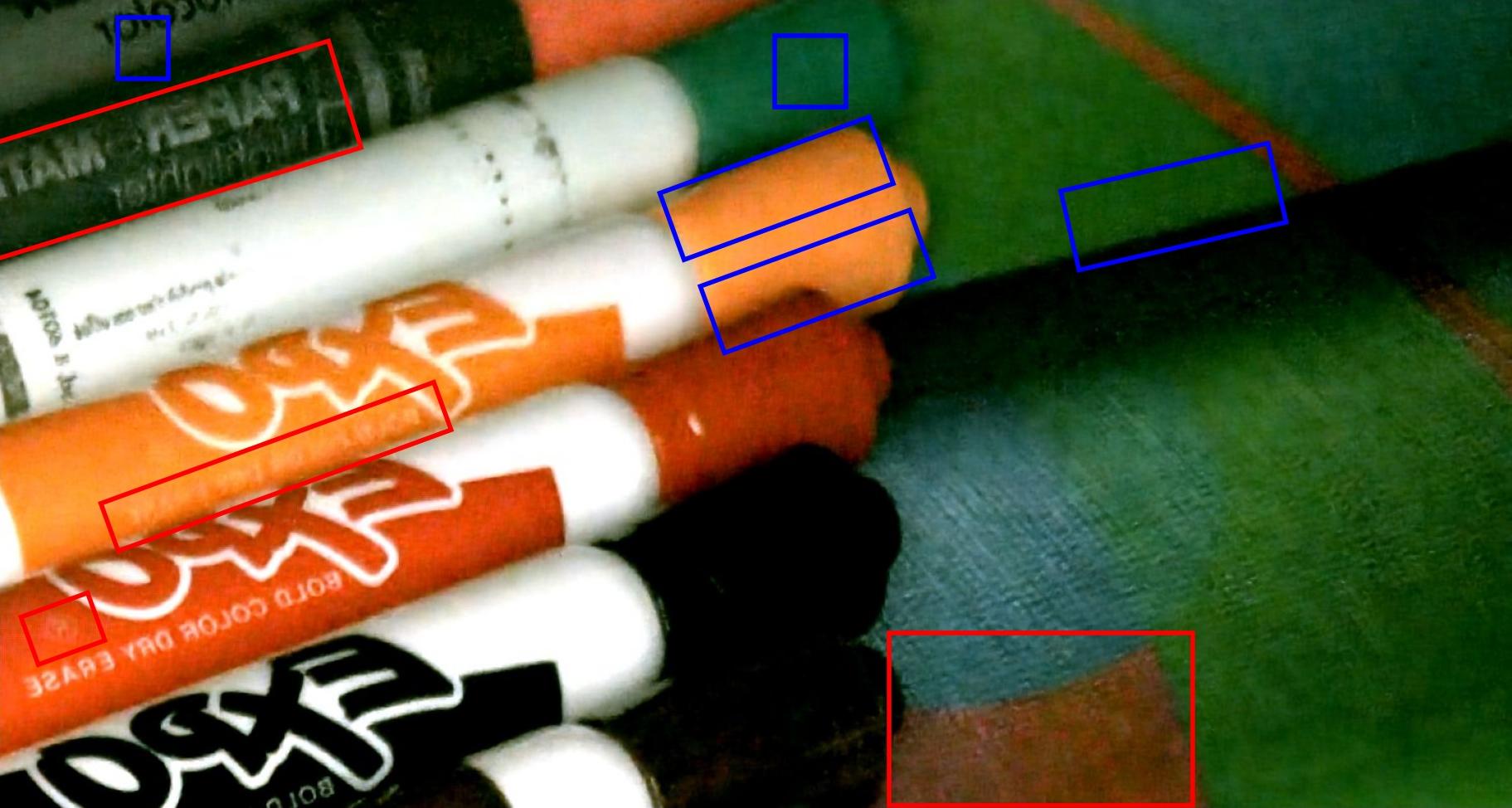}} \hfill
        \subfloat[self-supervised blind-spot]{\includegraphics[width=.249\textwidth]{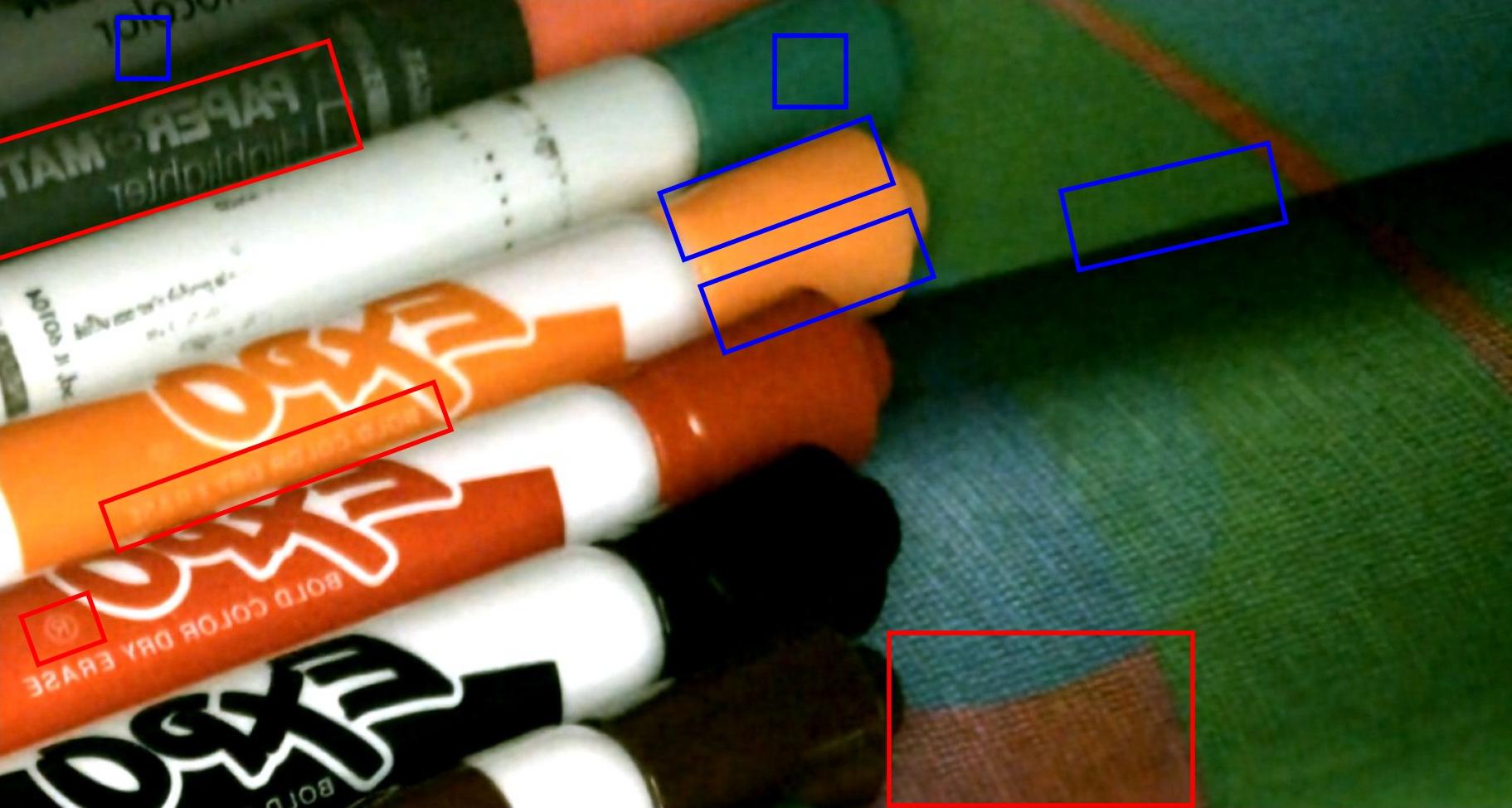}} \hfill
	\caption{Comparison of the different training strategies for the Experience II: the normal network trained with supervision (e) recovers less details than networks trained with self-supervision (see the \textcolor{red}{red} rectangles) and does not preserve correctly the colors (colors are washed in the orange and red pens). Furthermore, the model-supervised leaves some residual noise (see the \textcolor{blue}{blue} rectangles).}
	\label{fig:main_results_expII}
\end{figure*}

\begin{table}
\centering
\setlength{\tabcolsep}{0.68em}
\resizebox{\linewidth}{!}{
\begin{tabular}{ |c|c||c|c| }
 \hline
 \multirow{2}{*}{ISO} & Supervised & \multicolumn{2}{c|}{Self-supervised}  \\ \cline{2-4}
  & model-sup. normal & MF2F & blind-spot \\ \hline 
 3200 & 42.10 / .9865 & 43.63 / .9875 & 41.79 / .9840 \\
 12800 & 35.91 / 9681 & 38.80 / .9711 & 37.93 / .9688 \\
 \hline
\end{tabular}
}
\caption{Evaluation on the indoor CRVD dataset~\cite{yue2020supervised}.  No network was trained on this dataset (even the self-supervised ones).}
\label{tab:CRVD_indoors}
\end{table}

In this section, we first describe the setting of each of the three experiments together with the obtained results for both approaches. For better visualization, the video frames displayed in this section have been gamma corrected (with $\gamma = 2.2$), demosaicked with~\cite{MLRI} and white-balanced.

\subsection{Exp I: dynamic scenes with simulated noise}\label{sec:expI}


In this experiment we use the REDS 120 dataset to generate the clean raw data for both the \emph{surrogate dataset} and the \emph{synthetic dataset}. The difference lies on the noise \linebreak  model: we use the Poisson-Tukey lambda distribution of~\cite{wei2021physics} as noise model for the \emph{surrogate dataset}. This noise models extreme low-light conditions. In~\cite{wei2021physics} the authors provide parameters for three cameras. We use the noise parameters estimated for the Nikon D850. The noise in the \emph{synthetic dataset} is the heteroscedastic Gaussian with parameters set to approximate the Poisson-Tukey lambda noise of the \emph{surrogate dataset}.
All networks are pre-trained on the training split of the \emph{synthetic dataset}, and the fine-tunings are performed using the training split of the \emph{surrogate dataset}.

\paragraph{Results.} The first row of Table~\ref{tab:main_results} summarizes the average PSNR on our \emph{surrogate} validation set for the different training strategies. The results show that the self-supervised approaches outperform the supervised training in the \emph{synthetic dataset}: the self-supervised blind-spot network surpasses the model-supervised network by almost 0.7dB and has a much higher SSIM value. The results of the MF2F network have a PSNR similar to the model-supervised, but has a higher SSIM. 

From Figure~\ref{fig:main_results_expI}, we notice that both self-supervised networks recover more details and have a better reconstruction of the textures. The self-supervised blind-spot is even close to the gold standard. The result of MF2F has a small color shift, which is why it has a lower PSNR.
As the heteroscedastic Gaussian noise model does not fully approximate the noise of the \emph{surrogate real} test set, the model-supervised net results contain denoising artifacts which decrease its performance. On the contrary the self-supervised networks learn the actual noise of this simulated camera and produce results which compete with the gold standard.

\subsection{Exp II: real static videos as surrogate data}\label{sec:expII}

In the previous experiment, we use artificial ground truth in the \emph{surrogate dataset}. In this section, we are interested in the comparison between model-supervised and self-supervised on {real data}. To provide quantitative results, we use the \emph{Smartphone Image Denoising Dataset} (SIDD)~\cite{SIDD}, as it has ground truth. It provides images of ten static scenes, taken by five real cameras with different ISO levels, shutter speeds or illuminations levels. For each, the authors give an estimated ground-truth image obtained by averaging a burst of frames. We generate a ground truth constant video from the ground truth image, as there is no motion in the scene. We use eight static sequences of about 150 frames each obtained with the Google Pixel camera for ISO level 800. This \emph{surrogate dataset} is split into six sequences as a fine-tuning pool and two as a testing/validation pool.
For both supervised and self-supervised networks, quantitative results are evaluated on the testing pool of this \emph{surrogate dataset}. 

\begin{figure*}
    \centering
    \subfloat[gt]{\includegraphics[width=0.248\textwidth]{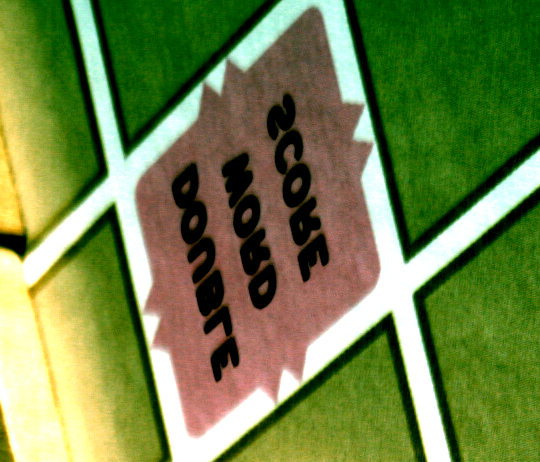}} \hfill
    \subfloat[real noisy]{\includegraphics[width=0.248\textwidth]{ 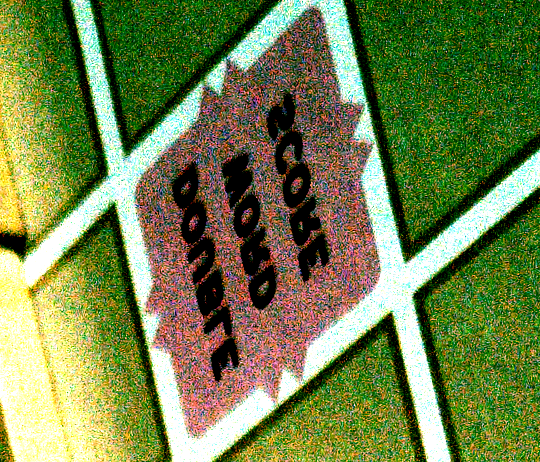}} \hfill
    \subfloat[Gold standard]{\includegraphics[width=0.248\textwidth]{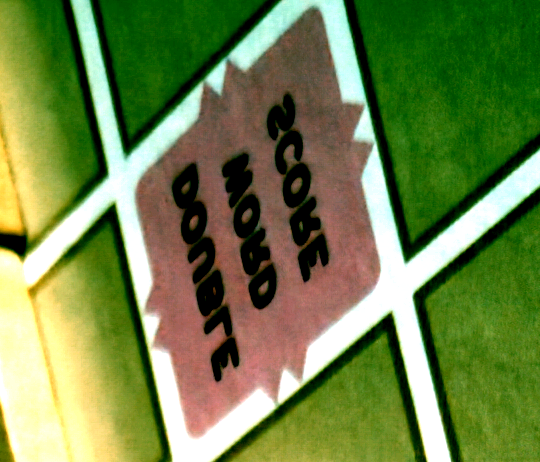}} \hfill
    \subfloat[Noise-ablation]{\includegraphics[width=0.248\textwidth]{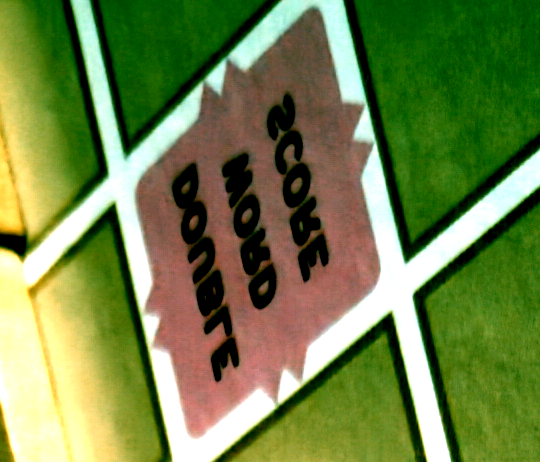}}\\
    \subfloat[Model-supervised normal]{\includegraphics[width=0.248\textwidth]{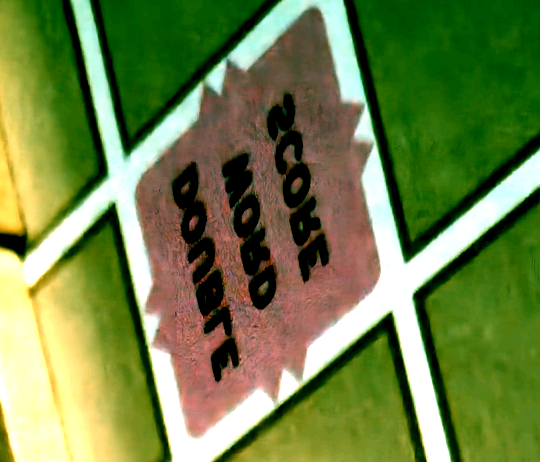}} \hfill
    \subfloat[Model-supervised blind-spot]{\includegraphics[width=0.248\textwidth]{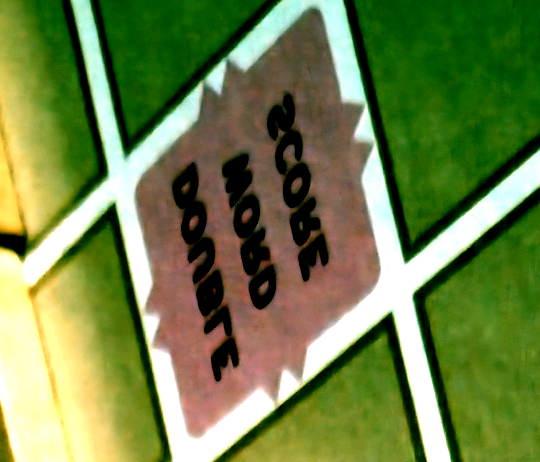}} \hfill
    \subfloat[MF2F]{\includegraphics[width=0.248\textwidth]{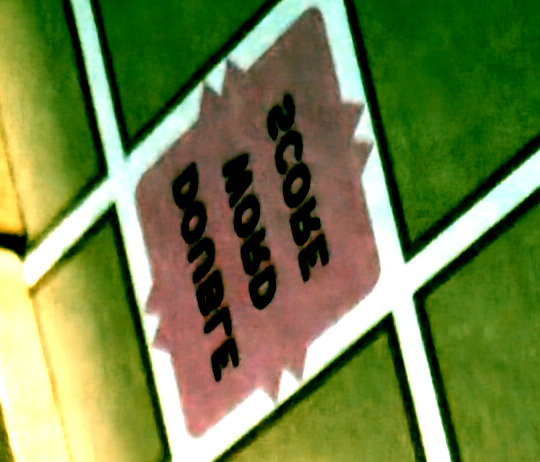}} \hfill
    \subfloat[Self-supervised blind-spot]{\includegraphics[width=0.248\textwidth]{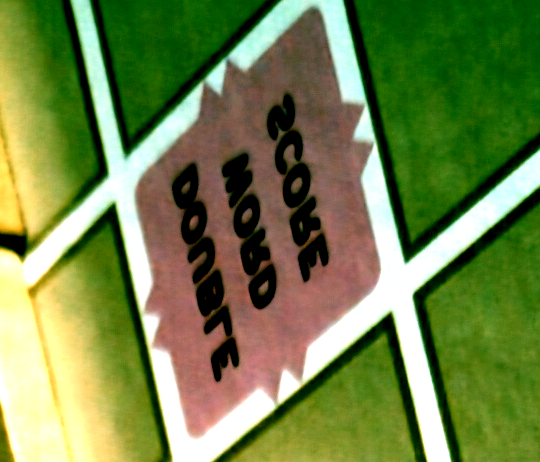}}
	\caption{Comparison of the different training strategies for the Experience II: the model-supervised normal leaves some residual noise (see around the letters).}
	\label{fig:artefacts_model_raw_expII}
\end{figure*}

\paragraph{Results.} The average PSNR and SSIM on the real validation set are presented in  Table~\ref{tab:main_results}. As for Experiment I, the trainings with self-supervision lead to a better performance than the trainings done in a supervised setting. On average, the self-supervised blind-spot outperforms the model-supervised by 1.5dB. 
In Figure~\ref{fig:main_results_expII}, the results with the self-supervised networks are sharper and have more details.
Figure~\ref{fig:artefacts_model_raw_expII} shows that the model-supervised network creates also artifacts (see near the text).

In this experiment, the \emph{synthetic dataset} differs from the \emph{surrogate} both in the noise model and the ground truth.  In order to differentiate this effects, we look at the results of the noise-ablation network, which is trained using clean real data as ground truth, with the simulated heteroscedastic noise. It is remarkable that the  result of the noise-ablation network matches exactly with the one of the gold standard (both in PSNR and SSIM). Visual inspection confirms that both results are indeed very similar. We deduce from this that in this case, the heteroscedastic noise model is a good approximation of the real noise, and therefore the problem of the model-supervised network is most likely due to the unprocessed synthetic clean data. On the contrary, in Experiment I, the clean data was the same for both the \emph{surrogate} and \emph{synthetic} datasets, thus the failure of model-supervised net was caused by a bad noise modeling.


\subsection{Exp III: real dynamic scenes}\label{sec:expIII}

In our final experiment, we will use the dataset introduced in \cite{yue2020supervised} for a visual comparison. It consists of \emph{real noisy raw videos of 10 outdoor dynamic scenes} acquired with a surveillance camera for five ISO levels. For such real data, we do not have ground-truth. 
As before we pre-train the networks on the \emph{synthetic} REDS 120 dataset with heteroscedastic noise.
We considered two ISO levels: 3200 and 12800, and fit the parameters of the noise model to approximate the real noise for each ISO level.

\vspace{-0.27cm}
\paragraph{Results.}
Visual results are shown in Figure~\ref{fig:main_results_expIII}. In this setting as well, the self-supervised training yields  more details leading to a better global reconstruction of the objects.

In~\cite{yue2020supervised}, the authors also acquire a dataset of videos with ground-truth of indoor scenes taken with the same camera (denoted CRVD). To simulate motion the authors produced stop-motion videos: the camera is fixed on a tripod and several images are taken for ground-truth estimation via averaging. Then, objects in the scene are slightly moved and the process is repeated to acquire new frames. This results in an unnatural motion. 
As an additional study, we evaluated the previous networks (trained on either the \emph{synthetic dataset} or the \emph{real} outdoor data with real motion) on this indoor dataset. No fine-tunings  were done to the indoor dataset as it is very small (10 sequences of only 7 frames each). In particular, the self-supervised networks were trained for the CRVD outdoor dataset and all the networks were trained for real motion. This study is another illustration of the network behavior in case of dataset bias. The quantitative results for two ISO levels 3200 and 12800 are gathered in Table~\ref{tab:CRVD_indoors}. For both ISOs, self-supervised outperforms the supervised network.  

%

\section{Conclusion}

In this work we propose a protocol to compare in fair conditions two training approaches for denoising real raw videos: supervised training on synthetic data and self-supervised training on the real data.
The difficulty of acquiring real videos with ground truth prevents us for doing a simple comparison. To address this issue, we set three experiments covering different use cases such as low light conditions, real motion, real noise at different ISO levels. 
In all cases, the self-supervised approaches outperformed the supervised one.
Among self-supervised techniques, the blind-spot approach UDVD gave better results than MF2F. The main caveat of UDVD is that blind-spot networks tend to be costlier. MF2F can be used to train any multi-frame network architecture.
Our experiments also shed light on how to improve the supervised approach. For normal illumination conditions (such as in the SIDD dataset) the main cause of the generalization gap of supervised training on synthetic data, is not necessarily the simple heteroscedastic Gaussian noise, indicating that more effort needs to be put in better modeling of the clean raw data.


\vspace{-0.5cm}
\paragraph{Acknowledgments.}
Work supported by a grant from MENRT. It was also partly financed by Office of Naval research grant N00014-17-1-2552. 
This work was performed using HPC resources 
from GENCI–IDRIS (grants 2022-AD011012453R1 and 2022-AD011012458R1) and  from the “Mésocentre” computing center of CentraleSupélec and ENS 
Paris-Saclay supported by CNRS and Région Île-de-France (http://mesocentre.centralesupelec.fr/).

{\small
\bibliographystyle{ieee_fullname}
\bibliography{my_bib,deep-denoising,image_denoising,video_denoising}
}

\end{document}


\title{[Supplementary material] Self-supervision versus synthetic datasets: which is the lesser evil in the context of video denoising?}

\author{Valéry Dewil \quad Arnaud Barral \quad Gabriele Facciolo \quad Pablo Arias\\
Université Paris-Saclay, CNRS, ENS Paris-Saclay, Centre Borelli, 91190, Gif-sur-Yvette, France\\
{\href{https://centreborelli.github.io/VDU2020-the-lesser-evil}{\tt\small https://centreborelli.github.io/VDU2020-the-lesser-evil}}
}

\maketitle

\section{Unprocessing of sRGB dataset}

Our \emph{synthetic dataset} is synthesized from the REDS dataset~\cite{reds}. It consists in clean sRGB videos with real motion. In order to create the \emph{synthetic dataset}, we need two steps: unprocess back sRGB data to the raw domain and adding realistic noise.
In this section, we talk about the unprocessing steps. We follow \emph{Brooks et al.}~\cite{brooks2019unprocessing}, with some modifications to adapt it to our case. First the REDS dataset is made with 8-bits quantized frames. To reduce the effect of the quantization we add a quantization noise to each pixel value sampled from uniform distribution in the range $[-1/2, 1/2]$. Originally, the authors of~\cite{brooks2019unprocessing} provide the \emph{Camera Color Matrix} for four different cameras from the \emph{Darmstadt Noise Dataset (DND)}~\cite{plotz2017benchmarking}. In our case we want to simulate a single camera, thus we use only one of them. 

The white balance is image dependent and thus inverting it is not straightforward. In~\cite{brooks2019unprocessing}, the authors estimated a range of realistic red and blue gains from \emph{DND} (normalized with respect to the green gain being set to 1). They found that the red gain $g^{\text{R}}$ has to be sampled uniformly in $[1.9, 2.4]$ and the blue gain $g^{\text{B}}$ in the range $[1.5, 1.9]$. They also consider a global gain $g^{\text{global}}$ applied to all channels (to invert the brightness adjustment in the forward pipeline). This global gain is sampled from a Gaussian distribution $\mathcal{N}(0.8, 0.1)$.
The total per-channel gain for channel $c$ is then $g^{\text{global}} / g^c$.
Occasionally the global gain can become greater than 1, which causes saturation later in the pipeline.
This is wanted by \emph{Brooks et al.} to create highlights and saturation. However, none of our \emph{surrogate datasets} contains saturated areas, thus we prevent our per-channel gain to exceed one by sampling a global gain from a truncated Gaussian instead, clipping its value to one.


For each experiment, the clean \emph{synthetic} raw dataset is tailored to model the \emph{surrogate dataset}. 
We use the same Bayer pattern and we match the ranges of both datasets. For that purpose, we apply to the synthetic videos an affine tone mapping that maps the $1\%$ and the $99\%$ percentiles of the \emph{synthetic} dataset to those of the \emph{surrogate dataset}.

The next subsection describes how we generate the noisy counterpart of the clean raw data.

\section{Simulating realistic noise}
Let $\{u_i\}_I$ be the set of unprocessed clean data and $\{\tilde{v}_j\}_J$ be a dataset of real noisy data (the \emph{surrogate dataset}). Given the clean data $\{u_i\}_I$ we can generate realistic noisy data $\{v_i\}_I$ by applying the heteroscedastic Gaussian noise model. For that purpose, the steps to follow are:

\smallskip\noindent (1) Estimate from $\{\tilde{v}_j\}_J$ the parameters $a$ and $b$ of an heteroscedastic Gaussian noise model. 

\smallskip\noindent (2) Simulate a set of sequences with synthetic noise $\{v_i\}_I$ where each $v_i = u_i + n_i$ with $n_i \sim \mathcal{N} \left( 0, \sqrt{a u_i + b} \right)$. The pairs of sequences  $\left( \{u_i\}_I, \{v_i\}_I \right)$  can then be used for training with supervision.   

For addressing the point 1, we used two different strategies. For the Experiment I, we model a camera with a synthetic noise generator~\cite{wei2021physics} and thus we can simulate the acquisition of flat-field images. Contrarily for the Experiment II and III, we want to model the noise model of a given camera having only a few noisy sequences. We followed two different methods to evaluate the noise model parameters. Both are described in the next subsection.

\nada{
In the end we obtain two networks one trained with supervision but with an approximated noise model  $\mathcal{F}_{\theta}^{Sup}$ and another trained with  self-supervision $\mathcal{F}_{\tilde{\theta}}^{SS}$ on the real data.

A part of the clean data $\{u_j\}_{j\in J \not\subset I}$ along with the corresponding realistic noisy data $\{ \tilde{v}_j \}_{j \in J}$ is used for validation. (Because the realistic noise model is assumed to be the reality). Both networks are evaluated on $\{ \tilde{v}_j \}_{j \in J}$ by computing PSNR and SSIM with respect to the clean data $\{u_j\}_{j\in J \not\subset I}$. 
}

\subsection{Noise parameters estimation}

\begin{figure}
    \centering
    \includegraphics[width=\linewidth]{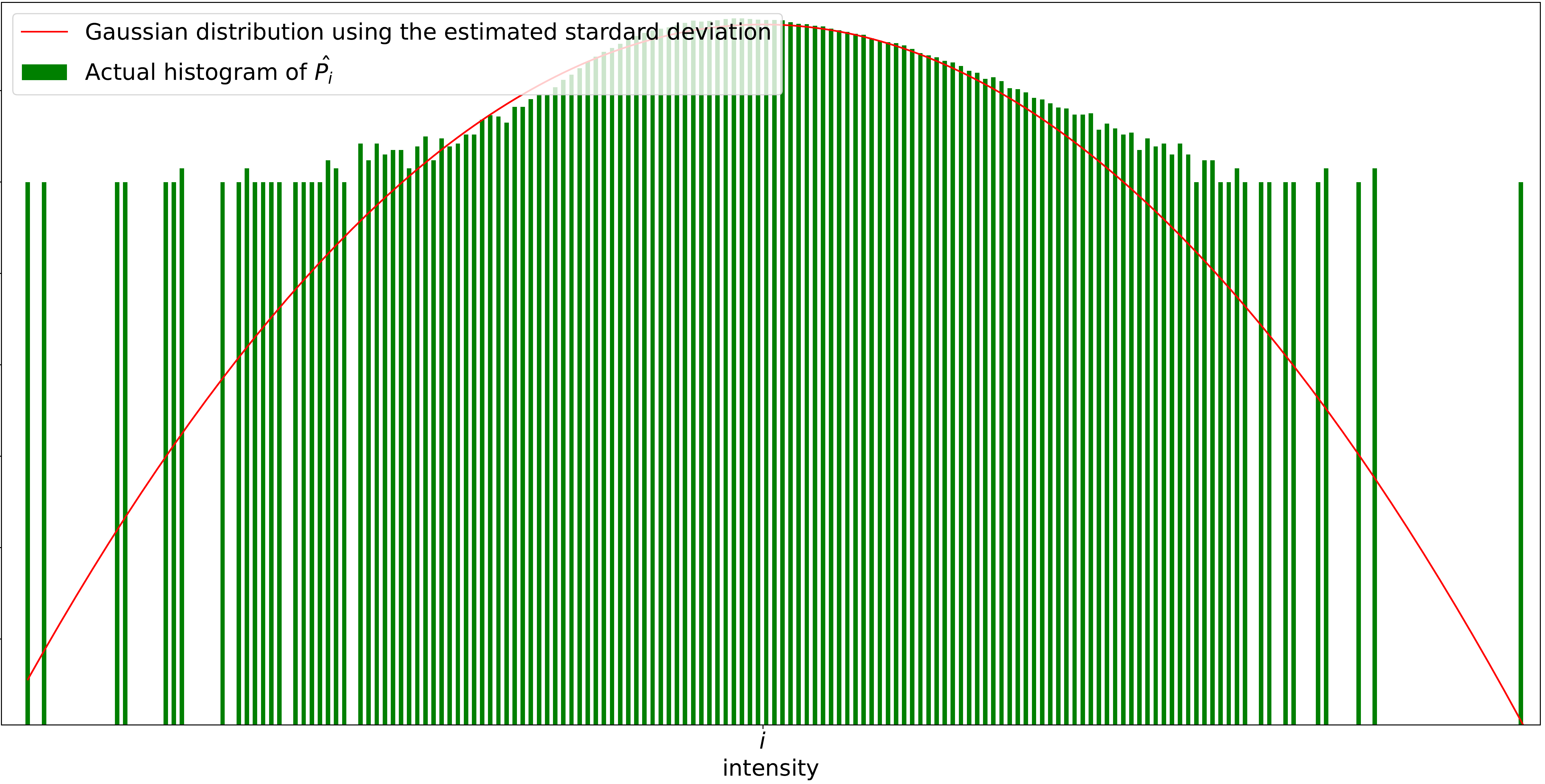}
    \caption{Example of actual histogram of the physic-based noise model (\cite{wei2021physics}) and the heteroscedastic Gaussian fitting (the \emph{y-axis} is in logarithmic scale).}
    \label{fig:histo_variance_TL_vs_gaussian}
\end{figure}

\begin{figure}
    \centering
    \includegraphics[width=\linewidth]{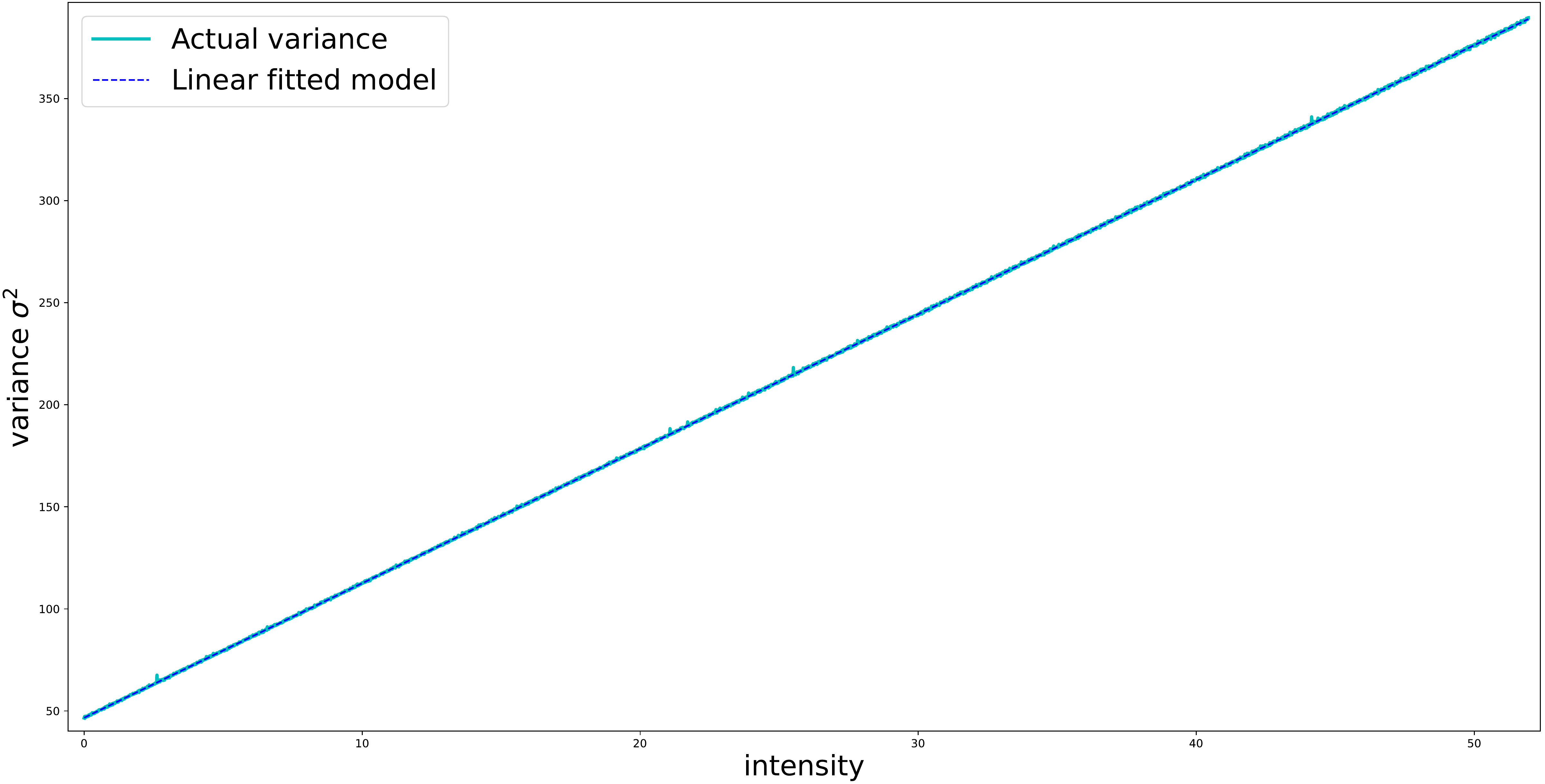}
    \caption{Variance of the noise model~\cite{wei2021physics} and the estimated linear model.}
    \label{fig:plot_variance_physic_model}
\end{figure}

\begin{figure}
    \centering
    \includegraphics[width=\linewidth]{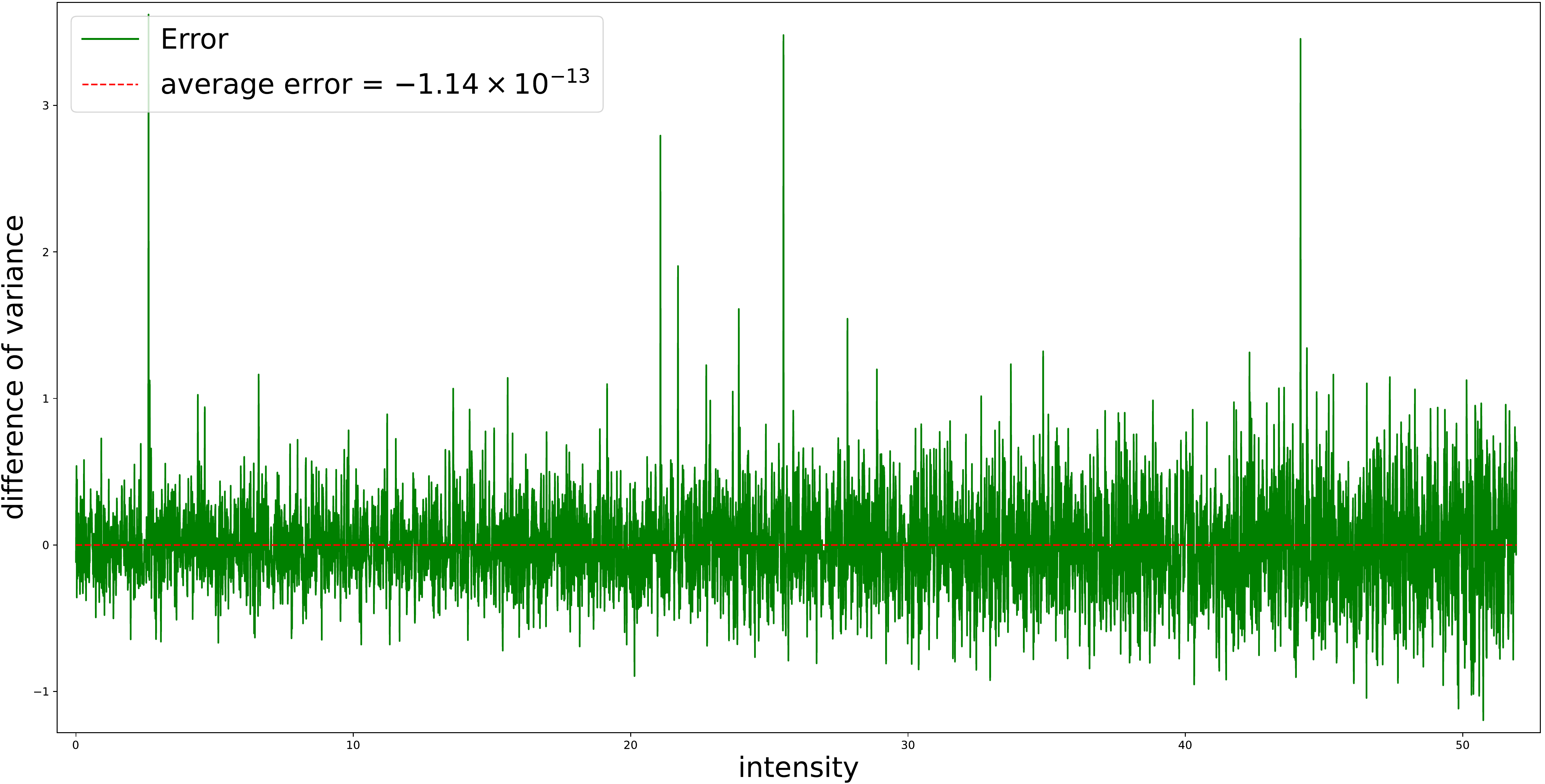}
    \caption{Difference of the variance of the noise model~\cite{wei2021physics} and the estimated linear model.}
    \label{fig:plot_variance_physic_model_error}
\end{figure}

\paragraph{Estimation for Experiment I.} In Experiment I we use the noise model introduced in~\cite{wei2021physics}, which models extreme low-light noise as a sum of a Poisson and a Tukey lambda distributions. In that sense, we have a simulated camera and the goal is to model its noise by a heteroscedastic Gaussian model whose variance is $\sigma^2(u) = a u + b$, where $u$ is the clean frame. 
To calibrate the $a,b$ parameters, we simulate the acquisition of \emph{flat-field} images, which is the usual way to calibrate signal dependent noise models.

We sample a range of constant patches $P_i$ with intensity level $i$. For each patch $P_i$ we generate a noisy patch $\hat{P}_i$ using the Poisson-Tukey lambda noise model and compute the variance $\sigma_i$ of the noisy patches. The parameters $a$ and $b$ are deduced from the points cloud $(i, \sigma_i)$ using the least square error method. Figure~\ref{fig:plot_variance_physic_model} shows a plot of this points cloud and the estimated linear model. 
The variance estimated from the Poisson-Tukey lambda noise is (as expected) an affine function of the intensity, therefore the affine model fits perfectly. Figure~\ref{fig:plot_variance_physic_model_error} shows the difference between the actual variance and the estimated linear model. We can see that the error is very small relatively to the variance value. The heteroscedastic Gaussian model will have the same intensity-variance curve, but the distributions are very different.
Figure~\ref{fig:histo_variance_TL_vs_gaussian} shows the histogram of the variance for a patch of middle range intensity. The estimated Gaussian distribution is also displayed. It can be seen that around the mean, the Poisson-Tukey lambda noise is well approximated by the Gaussian distribution. However, the Tukey lambda component has heavier tails than the Gaussian distribution. 

\begin{figure*}
    \centering
    \includegraphics[width=\textwidth]{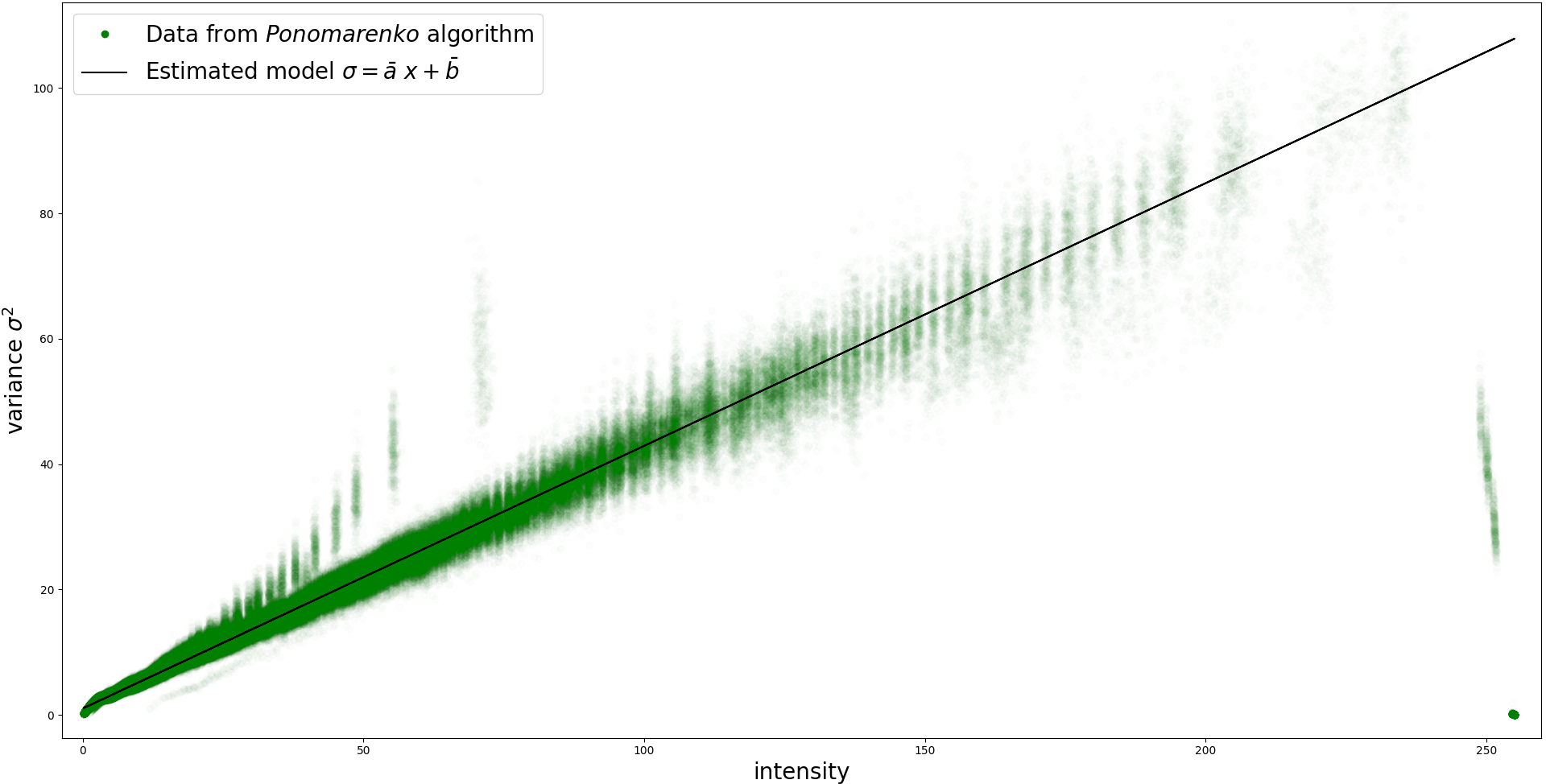}
    \caption{Linear model estimation of the noise curve.}
    \label{fig:ax+b_curve_exp_II}
\end{figure*}

\paragraph{Estimation in the Experiments II and III.} In the case of Experiments II and III, the \emph{surrogate} datasets consist of real noisy sequences but cannot generate more samples. Thus we need to estimate the camera noise level function (NLF) directly from the real data (SIDD \cite{SIDD} in Experiment II or CRVD~\cite{yue2020supervised} for the Experiment III). For that purpose, we estimate the NLF of each frame from each sequence of the \emph{surrogate dataset} using the method of \emph{Ponomarenko}~\cite{ponomarenko2007automatic,13-colom-ponomarenko-ipol}. For each individual noisy frame $v_i$, this method estimates a set of  intensity-variance points which are samples from the NLF.
We gather estimated intensity-variance points of each frame into a large point cloud. Figure~\ref{fig:ax+b_curve_exp_II} shows this point cloud for Experiment II (one camera of the SIDD dataset). We use transparent points, thus the level of opacity gives an indication of the density in the point cloud.
We then fit an affine model $\sigma^2(u) = a u + b$ by minimizing the least square error. 

{\small
\bibliographystyle{ieee_fullname}
\bibliography{my_bib,deep-denoising,image_denoising,video_denoising}
}